\documentclass[1p]{elsarticle}
\usepackage[english]{babel}
\usepackage[usenames,dvipsnames,svgnames,table]{xcolor}
\usepackage{graphicx}
\usepackage{epstopdf}

\usepackage[dvips]{epsfig}
\usepackage{booktabs}
\usepackage{amssymb}

\usepackage{ifthen}
\usepackage{fancyvrb}
\usepackage[dotinlabels]{titletoc}
\usepackage{array}
\usepackage{soul}
\usepackage{rotating}
\usepackage{psfrag}
\usepackage{url}
\usepackage{listings}
\usepackage{xspace}
\usepackage{subfigure}
\usepackage{multirow}

\usepackage{algorithm}
\usepackage{algorithmic}
\usepackage{tikz}

\definecolor{darkred}{rgb}{0.44,0,0}
\definecolor{darkgreen}{rgb}{0,0.44,0}
\definecolor{darkblue}{rgb}{0,0,0.44}
\definecolor{grey}{rgb}{0.5,0.5,0.5}

\definecolor{mygreen}{rgb}{0,0.6,0} \definecolor{mygray}{rgb}{0.5,0.5,0.5} \definecolor{mymauve}{rgb}{0.58,0,0.82}

\definecolor{dodgerblue}{rgb} {0.4,0.4,1.0} 
\definecolor{forestgreen}{rgb}{1.0,0.0,0.0} 
\definecolor{yellow1}{rgb}    {0.0,1.0,0.0} 
\definecolor{red}{rgb}        {1.0,0.0,1.0} 
\definecolor{lightgray}{rgb}{0.950,0.950,0.950}

\newcommand{\gemm}{{\sc gemm}\xspace}
\newcommand{\syrk}{{\sc syrk}\xspace}
\newcommand{\trsm}{{\sc trsm}\xspace}

\begin{document}
\begin{frontmatter}
\title{Revisiting Conventional Task Schedulers to Exploit Asymmetry in ARM big.LITTLE Architectures for Dense Linear Algebra}



\author[ucm]{Luis Costero}
\ead{lcostero@ucm.es}
\author[ucm]{Francisco D. Igual\corref{cor1}}
\ead{figual@ucm.es}
\author[ucm]{Katzalin Olcoz}
\ead{katzalin@ucm.es}
\author[uji]{Enrique~S.~Quintana-Ort\'{\i}}
\ead{quintana@ucm.es}

\cortext[cor1]{Corresponding author}

\address[ucm]{Departamento de Arquitectura de Computadores y Autom\'atica, Universidad Complutense de Madrid, Madrid, Spain}
\address[uji]{Departamento de Ingenier\'{\i}a y Ciencia de Computadores, Universidad Jaume I, Castell\'on, Spain}


\begin{abstract}
Dealing with asymmetry in the architecture opens a plethora of questions from the perspective of scheduling task-parallel 
applications,  and there exist early attempts to address this problem via {\em ad-hoc} strategies embedded into a
runtime framework. In this paper we take a different path, which consists in addressing
the complexity of the problem at the library level, via a few asymmetry-aware fundamental kernels, 
hiding the architecture heterogeneity from the task scheduler.
For the specific domain of dense linear algebra, we show that this is not only possible but delivers much higher performance
than a naive approach based on an asymmetry-oblivious scheduler. Furthermore, this solution also outperforms
an {\em ad-hoc} asymmetry-aware scheduler furnished with sophisticated scheduling techniques.

\end{abstract}


\end{frontmatter}

\section{Introduction}
\label{sec:introduction}

The end of Dennard scaling 
has promoted heterogeneous systems into a mainstream approach to
leverage the steady growth of transistors on chip dictated by Moore's law~\cite{Esm11,Dur13}.
ARM\textregistered\ big.LITTLE\texttrademark\ processors are a particular class of heterogeneous architectures that
combine two types of multicore clusters, consisting of a few high performance (big) cores and a collection of low power (LITTLE) cores.
These {\em asymmetric multicore processors} (AMPs)
can in theory deliver much higher performance for the same power budget.
Furthermore, 
compared with multicore servers equipped with graphics processing units (GPUs), NVIDIA's Tegra chips
and AMD's APUs, ARM big.LITTLE processors differ in that the cores in these systems-on-chip (SoC) share the 
same instruction set architecture and a strongly coupled memory subsystem.

Task parallelism has been reported as an efficient means to tackle the considerable number of cores 
in current processors. Several efforts, pioneered by Cilk~\cite{cilkweb}, aim to ease the development
and improve the performance of task-parallel programs by embedding task scheduling inside a 
{\em runtime} (framework). The benefits of this approach for complex dense linear algebra (DLA) operations
have been demonstrated, among others, by OmpSs~\cite{ompssweb}, StarPU~\cite{starpuweb}, PLASMA/MAGMA~\cite{plasmaweb,magmaweb},
Kaapi~\cite{kaapiweb}, and {\tt libflame}~\cite{flameweb}.
In general, the runtimes underlying these tools decompose DLA routines into a collection
of numerical kernels (or tasks), 
and then take into account the dependencies between the tasks in order to correctly issue their execution to the system cores. 
The tasks are therefore the ``indivisible'' scheduling unit while the cores constitute the basic computational resources.

In this paper we introduce an efficient approach to execute task parallel DLA routines on AMPs via conventional asymmetry-oblivious schedulers.
Our conceptual solution aggregates the cores of the AMP into a number of 
{\em symmetric virtual cores} (VCs) which become the only basic computational resources that are visible to the runtime scheduler. 
In addition, an specialized implementation of each type of task, from an asymmetry-aware DLA library, partitions each numerical kernel into 
a series of finer-grain computations, which are efficiently executed by the asymmetric aggregation of cores of a single VC. 
Our work thus makes the following specific contributions:
\begin{itemize}
\item We target in our experiments the Cholesky factorization~\cite{GVL3}, a complex operation for the solution of symmetric
      positive definite linear systems that is representative of many other DLA computations.
      Therefore, we are confident that our approach and results carry beyond a significant fraction of 
      LAPACK ({\em Linear Algebra PACKage})~\cite{lapack}.
\item For this particular factorization, we describe how to leverage the asymmetry-oblivious task-parallel runtime scheduler in OmpSs, 
      in combination with a data-parallel instance of the BLAS-3 ({\em basic linear algebra subprograms})~\cite{blas3}
      in the BLIS library specifically designed for ARM big.LITTLE AMPs~\cite{asymBLIS,BLIS1}.
\item We provide practical evidence that, compared with an {\em ad-hoc} asymmetry-conscious scheduler recently developed 
      for OmpSs~\cite{OmpSsbigLITTLE}, 
      our solution yields higher performance for the multi-threaded execution of the Cholesky factorization on 
      an Exynos 5422 SoC comprising two quad-core clusters, with ARM Cortex-A15 and Cortex-7 technology.
\item In conclusion, compared with previous work~\cite{asymBLIS,OmpSsbigLITTLE}, this paper demonstrates that, 
      for the particular domain of DLA, it is possible to hide the difficulties intrinsic to dealing with an asymmetric architecture
      (e.g., workload balancing for performance, energy-aware mapping of tasks to cores, and criticality-aware scheduling)
      inside an asymmetry-aware implementation of the BLAS-3. As a consequence,
      our solution can refactor any conventional (asymmetry-agnostic) scheduler to exploit the task parallelism present in complex DLA operations.
\end{itemize}

The rest of the paper is structured as follows. 
Section~\ref{sec:architecture} presents the target ARM big.LITTLE AMP, together with the main execution paradigms it exposes.
Section~\ref{sec:related} reviews the state-of-the-art in runtime-based task scheduling
and DLA libraries for (heterogeneous and) asymmetric architectures.
Section~\ref{sec:runtime} introduces the approach to reuse conventional runtime task
schedulers on AMPs by relying on an asymmetric DLA library.
Section~\ref{sec:results} reports the performance results attained by the
proposed approach; and Section~\ref{sec:conclusions} closes the paper with the
final remarks.



\section{Software Execution Models for ARM big.LITTLE SoCs}
\label{sec:architecture}

The target architecture for our design and evaluation is an ODROID-XU3 board comprising a Samsung Exynos 5422 SoC with an ARM
Cortex-A15 quad-core processing cluster (running at 1.3~GHz) and a Cortex-A7 quad-core processing 
cluster (also operating at 1.3~GHz).  Both clusters access a shared DDR3 RAM (2~Gbytes) via 128-bit coherent bus interfaces.
Each ARM core (either Cortex-A15 or Cortex-A7) has a 32+32-Kbyte L1 (instruction+data) cache.
The four ARM Cortex-A15 cores share a 2-Mbyte L2 cache, while the four ARM Cortex-A7 cores
share a smaller 512-Kbyte L2 cache.

Modern big.LITTLE SoCs, such as the Exynos 5422, 
offer a number of software execution models with support from the operating system (OS):
\begin{enumerate}
 \item {\em Cluster Switching Mode (CSM)}: The processor is logically divided into two clusters, one containing the big cores 
         and the other with the LITTLE cores, but only one cluster is usable at any given time. 
         The OS transparently activates/deactivates the clusters depending on the workload in order to balance
	 performance and energy efficiency.
 \item {\em CPU migration (CPUM)}: The physical cores are grouped into pairs, 
         each consisting of a fast core and a slow core, 
         building VCs to which the OS maps threads. At any given moment, only one physical core is active per VC,
         depending on the requirements of the workload. In big.LITTLE specifications where the number of fast and slow cores do
         not match, the VC can be assembled from a different number of cores of each type. 
         The {\em In-Kernel Switcher} (IKS) is Linaro's solution for this model.
 \item {\em Global Task Scheduling (GTS)}. This is the most flexible model. 
         All 8 cores are available for thread scheduling, and the OS maps the threads 
	 to any of them depending on the specific nature of the workload and core availability. 
         ARM's implementation of GTS is referred to as big.LITTLE MP.
\end{enumerate}

\begin{figure}[tbh!]
 \centering
 \subfigure[CSM]{
   \includegraphics[width=0.4\textwidth]{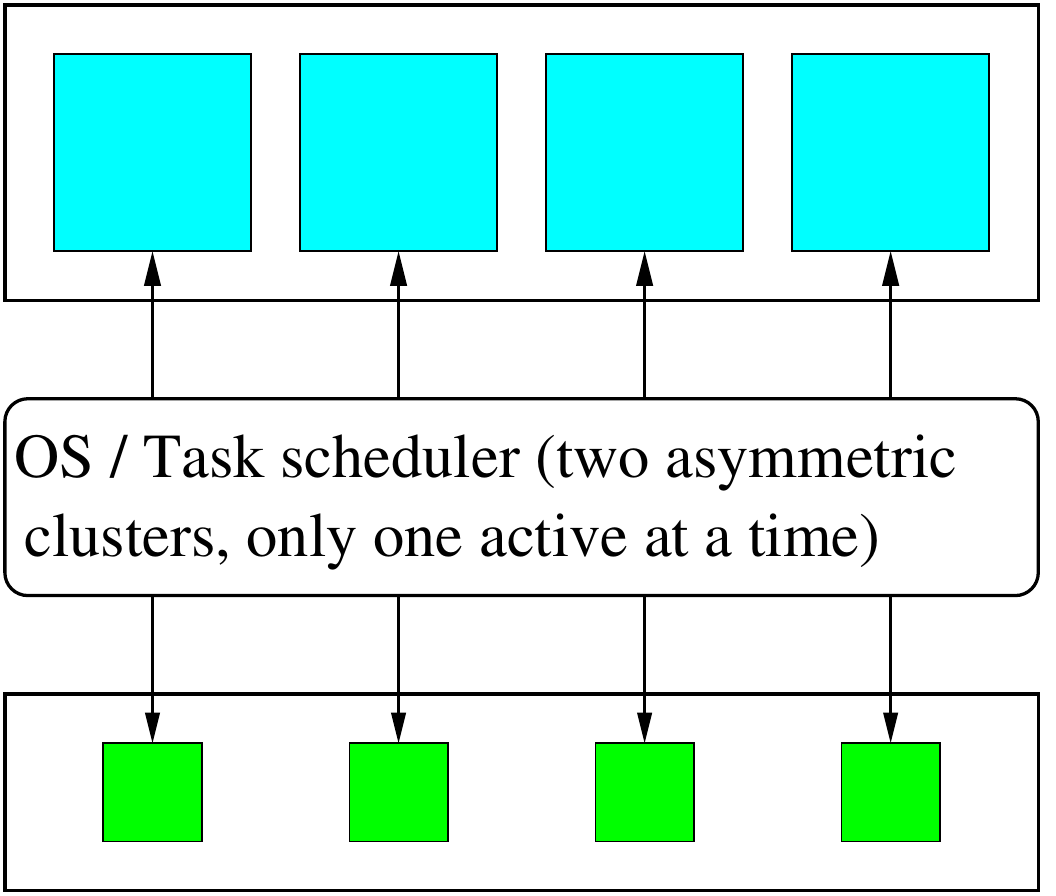}
   }
 \subfigure[CPUM]{
   \includegraphics[width=0.45\textwidth]{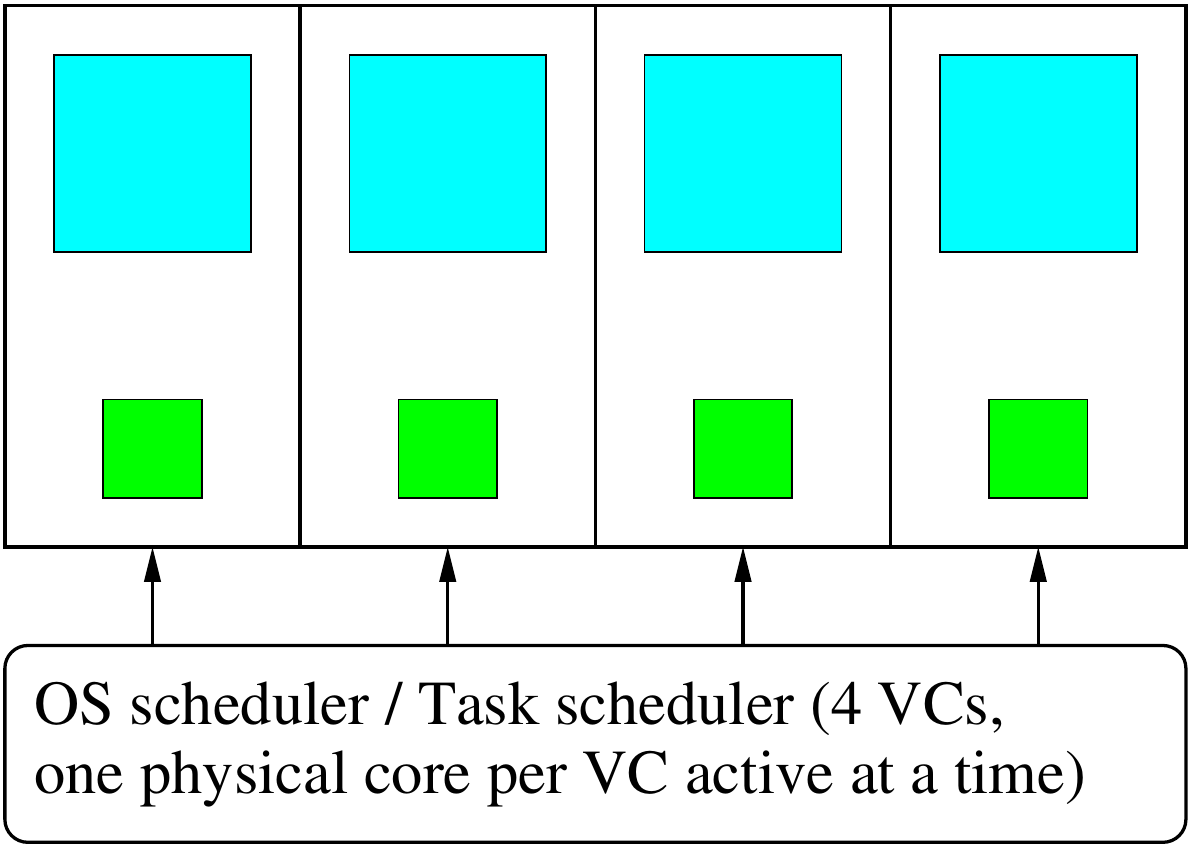}
   }
 \subfigure[GTS]{
   \includegraphics[width=0.6\textwidth]{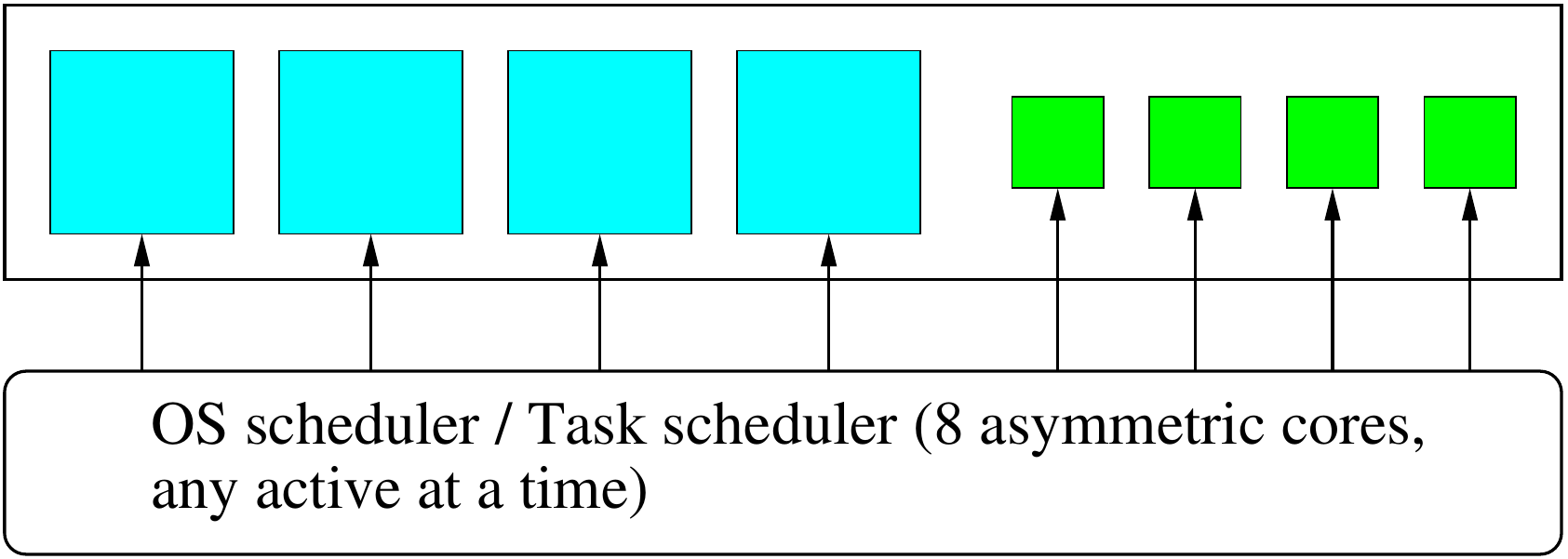}
   }
   \caption{Operation modes for modern big.LITTLE architectures.}
   \label{fig:modes}
\end{figure}

Figure~\ref{fig:modes} offers an schematic view of these three execution models for modern big.LITTLE architectures. GTS is the most
flexible solution, allowing the OS scheduler to map threads to any available core or group of cores. 
GTS exposes the complete pool of 8 (fast and slow) cores in the Exynos 5422 SoC
to the OS. This allows a straight-forward port of existing
threaded application, including runtime task schedulers, to exploit all the computational resources
in this AMP, provided the multi-threading technology underlying the software is based on conventional
tools such as, e.g., POSIX threads or OpenMP. Attaining high performance in asymmetric architectures,
even with a GTS configuration, is not as trivial, and is one of the goals of this paper.

Alternatively, CPUM proposes a pseudo-symmetric
view of the Exynos 5422, 
transforming this 8-core asymmetric SoC into 4~symmetric multicore processors (SMPs), which
are logically exposed to the OS scheduler.
(In fact, as this model only allows one active core per VC, but the type of the specific 
core that is in operation can differ from one VC to another, the CPUM is still asymmetric.) 

In practice, runtime task schedulers can mimic or approximate any of these OS operation modes. A straight-forward model is simply obtained by
following the principles governing GTS to map ready tasks to any available core. With this solution, 
load unbalance can be tackled via {\em ad-hoc} (i.e., asymmetry-aware) scheduling policies embedded into the runtime
that map tasks to the most ``appropriate'' resource. 


\section{Parallel Execution of DLA Operations on Multi-threaded Architectures}
\label{sec:related}

In this section we briefly review several software efforts, in the form of task-parallel runtimes
and libraries, that were specifically designed for DLA, or have been successfully applied in this domain,
{\em when the target is (an heterogeneous system or) an AMP}.

\subsection{Runtime task scheduling of complex DLA operations}

\subsubsection{Task scheduling for the Cholesky factorization}

We start by describing how to extract task parallelism during the execution of a DLA operation, 
using the Cholesky factorization as a workhorse example. This particular operation, which is representative of several other
factorizations for the solution of linear systems,
decomposes an $n \times n$ symmetric positive definite matrix $A$ into the product $A=U^TU$,
where the $n \times n$ Cholesky factor $U$ is upper triangular~\cite{GVL3}. 

\begin{lstlisting}[float=th!,frame=lines,caption=C implementation of the blocked Cholesky factorization.,label=lst:chol]
void cholesky (double *A[s][s], int b, int s)
{
   for (int k = 0; k < s; k++) {

      po_cholesky (A[k][k], b, b);              // Cholesky factorization
                                                // (diagonal block)
      for (int j = k + 1; j < s; j++) 
         tr_solve (A[k][k], A[k][j], b, b);     // Triangular solve

      for (int i = k + 1; i < s; i++) {
         for (int j = i + 1; j < s; j++)
            ge_multiply (A[k][i], A[k][j], 
                         A[i][j], b, b);        // Matrix multiplication
         sy_update (A[k][i], A[i][i], b, b);    // Symmetric rank-b update
      }

   }
}
\end{lstlisting}

\begin{figure}[tbh!]
\begin{center}
\includegraphics[scale=0.35]{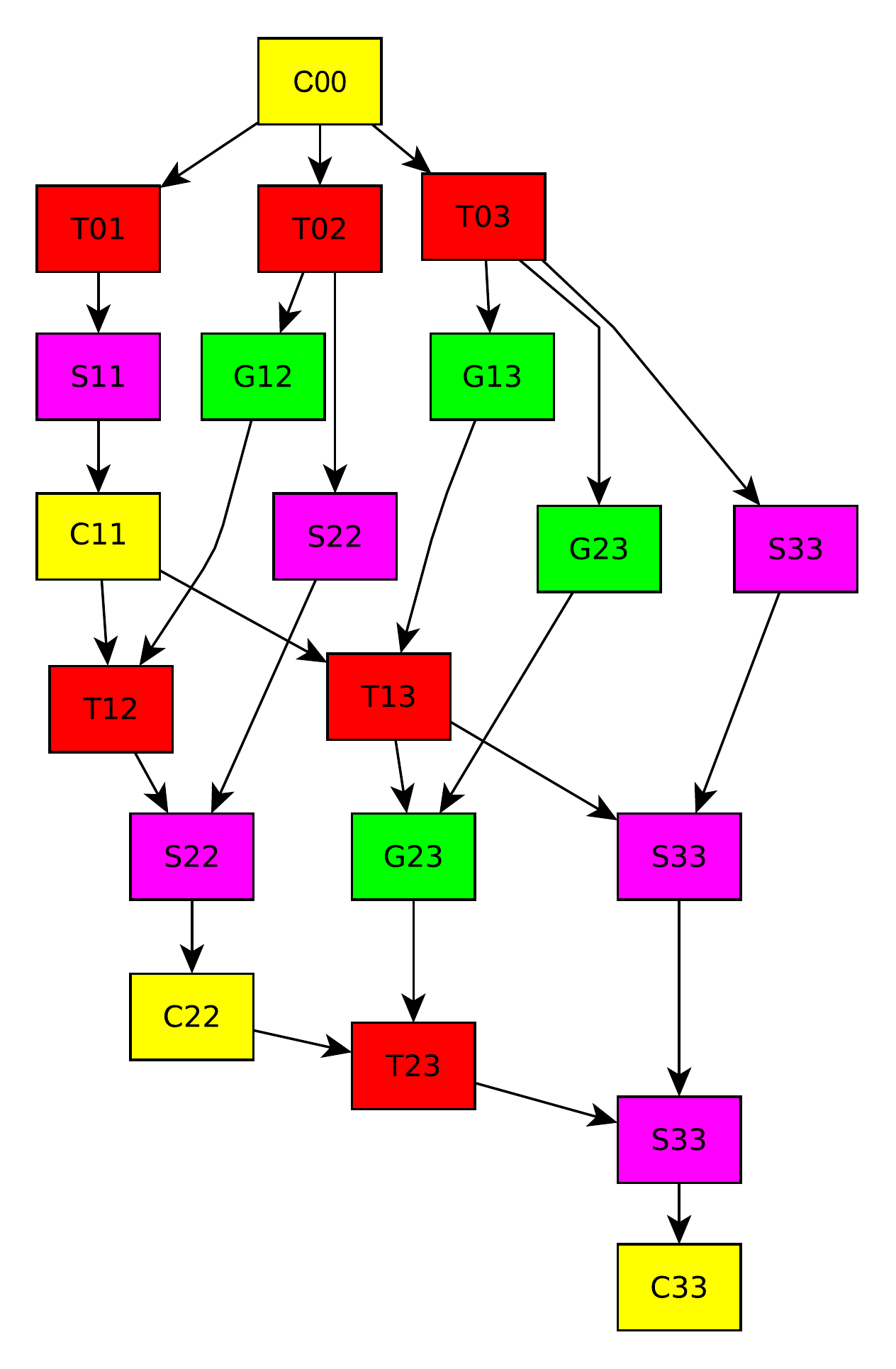}
\end{center}
\caption{DAG with the tasks and data dependencies resulting from the application of the code in Listing~\ref{lst:chol}
         to a $4 \times 4$ blocked matrix ({\tt s}=4). The labels specify the type of kernel/task
         with the following correspondence: 
         ``{\sf C}'' for the Cholesky factorization,
         ``{\sf T}'' for the triangular solve,
         ``{\sf G}'' for the matrix multiplication, and
         ``{\sf S}'' for the symmetric rank-{\tt b} update. 
         The subindices (starting at 0) specify the sub-matrix that the corresponding task updates, and the color is used
         to distinguish between different values of the iteration index {\tt k}.}
\label{fig:dag}
\end{figure}

Listing~\ref{lst:chol} displays a simplified C code for the factorization
of an {\tt n}$\times${\tt n} matrix {\tt A}, stored as {\tt s}$\times${\tt s} (data) sub-matrices 
of dimension {\tt b}$\times${\tt b} each.
This blocked routine 
decomposes the operation into a collection of building {\em kernels}:
{\tt po\_cholesky} (Cholesky factorization), {\tt tr\_solve} (triangular solve),
{\tt ge\_multiply} (matrix multiplication), and
{\tt sy\_update} (symmetric rank-{\tt b} update). 
The order in which these kernels are invoked during the execution of the routine, and the 
sub-matrices that each kernel read/writes,
result in a direct acyclic graph (DAG) that reflects the dependencies between tasks (i.e., instances of the kernels) and, therefore,
the task parallelism of the operation.
For example, Figure~\ref{fig:dag} shows the DAG with
the tasks (nodes) and data dependencies (arcs) intrinsic to the execution
of Listing~\ref{lst:chol}, when applied to a matrix composed of $4 \times 4$ sub-matrices (i.e., {\tt s}=4).  

The DAG associated with an algorithm/routine is a graphical representation of the task parallelism of the
corresponding operation, and a runtime system can exploit
this information to determine the task schedules that satisfy the DAG dependencies.
For this purpose, in OmpSs the programmer employs OpenMP-like directives ({\tt pragma}s) to annotate routines appearing in the
code as tasks, indicating the directionality
of their operands (input, output or input/output) by means of clauses.
The OmpSs runtime then decomposes the code (transformed by Mercurium source-to-source compiler) into a number of tasks at run time,
dynamically identifying dependencies among these, and issuing {\em ready tasks}
(those with all dependencies satisfied) for their execution to the processor cores of the system.

Listing~\ref{lst:chol_tasks} shows the annotations a programmer needs to add in order to exploit task parallelism using
OmpSs; see in particular the lines labelled with `{\tt \#pragma omp}''.
The clauses {\tt in}, {\tt out} and {\tt inout} denote data directionality, and help the task scheduler to keep track of 
data dependencies between tasks during the execution.
We note that, in this implementation, the four kernels simply boil down to calls to
four {\em fundamental computational kernels} for DLA from LAPACK 
({\tt dpotrf}) and the BLAS-3 ({\tt dtrsm}, {\tt dgemm} and {\tt dsyrk}).

\begin{lstlisting}[float=th!,frame=lines,caption=Labeled tasks involved in the blocked Cholesky factorization.,label=lst:chol_tasks]
#pragma omp task inout([b][b]A)
void po_cholesky (double *A, int b, int ld)
{
 static int        INFO = 0;
 static const char UP   = 'U';
 dpotrf (&UP, &b, A, &ld, &INFO);  // LAPACK Cholesky factorization
}

#pragma omp task in([b][b]A) inout([b][b]B)
void tr_solve (double *A, double *B, int b, int ld)
{
 static double     DONE = 1.0;
 static const char LE   = 'L', UP = 'U', TR = 'T', NU = 'N';
 dtrsm (&LE, &UP, &TR, &NU, &b, &b, 
        &DONE, A, &ld, B, &ld);    // BLAS-3 triangular solve
}

#pragma omp task in([b][b]A, [b][b]B) inout([b][b]C)
void ge_multiply (double *A, double *B, double *C, int b, int ld)
{
 static double     DONE = 1.0, DMONE = -1.0;
 static const char TR   = 'T', NT    = 'N';
 dgemm (&TR, &NT, &b, &b, &b, 
        &DMONE, A, &ld, B, &ld, 
        &DONE,  C, &ld);           // BLAS-3 matrix multiplication
}

#pragma omp task in([b][b]A) inout([b][b]C)
void sy_update (double *A, double *C, int b, int ld)
{
 static double     DONE = 1.0, DMONE = -1.0;
 static const char UP   = 'U', TR    = 'T';
 dsyrk (&UP, &TR, &b, &b, 
        &DMONE, A, &ld, 
        &DONE,  C, &ld);           // BLAS-3 symmetric rank-b update
}

\end{lstlisting}

\subsubsection{Task scheduling in heterogeneous and asymmetric architectures}
\label{subsec:ompss}

For servers equipped with one or more graphics accelerators, (specialized versions of) the schedulers underlying OmpSs,
StarPU, MAGMA, Kaapi and {\tt libflame} distinguish between the execution target being either a general-purpose core (CPU) or a GPU, 
assigning tasks to each type of resource depending on their properties, and
applying techniques such as data caching or locality-aware task mapping; see, 
among many others,~\cite{Quintana:2008:PMA,CPE:CPE1463,Augonnet:2011:SUP:1951453.1951454,5470941,Gautier:2013:XRS:2510661.2511383}.

The designers/developers
of the OmpSs programming model and the Nanos++ runtime task scheduler recently introduced a new version of their framework,
hereafter referred to as Botlev-OmpSs,
specifically tailored for AMPs~\cite{OmpSsbigLITTLE}. 
This asymmetry-conscious runtime embeds a scheduling policy
CATS (Criticality-Aware Task Scheduler) that relies on bottom-level longest-path priorities,
keeps track of the criticality of the individual tasks, and leverages this information, at execution time, to 
assign a ready task to either a critical or a non-critical queue. 
In this solution, tasks enqueued in the critical queue can only be executed by the fast cores.
In addition, the enhanced scheduler integrates uni- or bi-directional work stealing between fast and slow cores.
According to the authors, this sophisticated {\em ad-hoc} scheduling strategy for heterogeneous/asymmetric processors attains remarkable performance
improvements in a number of target applications; see~\cite{OmpSsbigLITTLE} for further details.

When applied to a task-parallel DLA routine, the asymmetry-aware scheduler in Botlev-OmpSs
maps each task to a single (big or LITTLE) core, and simply invokes a sequential DLA library to conduct the actual work.
On the other hand, we note that this approach required an important redesign of the underlying scheduling policy (and thus, a considerable
programming effort for the runtime developer), in order to exploit the heterogeneous architecture.
In particular, detecting the criticality of a task at execution time is a nontrivial question.


\subsection{Data-parallel libraries of fundamental (BLAS-3) DLA kernels}

\subsubsection{Multi-threaded implementation of the BLAS-3}

An alternative to the runtime-based (i.e., task-parallel) approach to execute 
DLA operations on multi-threaded architectures consists in relying on
a library of specialized kernels that statically partitions the work among the computational resources, 
or leverages a simple schedule mechanism such as those available, e.g.,
in OpenMP. For DLA operations with few and/or simple data dependencies, as is the case of the BLAS-3, 
and/or when the number of cores in the target
architecture is small, this option can avoid the costly 
overhead of a sophisticated task scheduler, providing a more efficient solution.
Currently this is preferred option for all high performance implementations of the BLAS for multicore processors,
being adopted in both commercial and open source packages such as, e.g., 
AMD ACML, IBM ESSL, Intel MKL, GotoBLAS~\cite{Goto:2008:AHP}, OpenBLAS~\cite{OpenBLAS} and BLIS.


BLIS in particular mimics GotoBLAS to orchestrate all BLAS-3 kernels (including the matrix multiplication, \gemm) as three nested loops
around two packing routines, which accommodate the data in the higher levels of the cache hierarchy,
and a {\em macro-kernel} in charge of performing the actual computations.
Internally, BLIS implements the macro-kernel as two additional loops around a {\em micro-kernel} that, in turn,
boils down to a loop around a symmetric rank-1 update. For the purpose of the following discussion, we will only consider the three outermost loops 
in the BLIS implementation of \gemm for the multiplication
$C:=C+A\cdot B$, where $A,B,C$ are respectively $m \times k$, $k\times n$ and $m \times n$ matrices, stored
in arrays {\tt A}, {\tt B} and {\tt C}; see Listing~\ref{lst:gemm}.
In the code, {\tt mc}, {\tt nc}, {\tt kc} are cache configuration parameters that need to be adjusted for performance taking
into account, among others, the latency of the floating-point units, number of vector
registers, and size/associativity degree of the cache levels.

\begin{lstlisting}[float=th!,frame=lines,caption=High performance implementation of \gemm in BLIS.,label=lst:gemm]
void gemm (double A[m][k], double B[k][n], double C[m][n], 
           int m,  int n,  int k, int mc, int nc, int kc)
{
   double *Ac = malloc (mc * kc * sizeof (double)), 
          *Bc = malloc (kc * nc * sizeof (double));

   for (int jc = 0; jc < n; jc+=nc) {              // Loop 1
      int jb = min(n-jc+1, nc);

      for (int pc = 0; pc < k; jc+=kc) {           // Loop 2
         int pb = min(k-pc+1, kc);

         pack_buffB (B[pc][jc], Bc, kb, nb);       // Pack A->Ac

         for (int ic = 0; ic < m; ic+=mc) {        // Loop 3
            int ib = min(m-ic+1, mc);

            pack_buffA (A[ic][pc], Ac, mb, kb);    // Pack A->Ac

            gemm_kernel (Ac, Bc, C[ic][jc], 
                         mb, nb, kb, mc, nc, kc);  // Macro-kernel
         }
      }
   }
}
\end{lstlisting}

\subsubsection{Data-parallel libraries for asymmetric architectures}

The implementation of \gemm in BLIS has been demonstrated to deliver high performance on a wide
range of multicore and many-core SMPs~\cite{BLIS2,BLIS3}. These studies offered a few relevant
insights that guided the parallelization of \gemm (and also other Level-3 BLAS) on the ARM big.LITTLE architecture under the GTS software execution model.
Concretely, the architecture-aware 
multi-threaded parallelization of \gemm in~\cite{asymBLIS} integrates the following three techniques:
\begin{itemize}
\item A dynamic 1-D partitioning of the iteration space to distribute the workload in
      either Loop~1 or Loop~3 of BLIS \gemm between the two clusters.
\item A static 1-D partitioning of the iteration space that distributes the workload of one of the loops 
      internal to the macro-kernel between the cores of the same cluster.
\item A modification of the control tree that governs the multi-threaded parallelization of BLIS \gemm in order to
      accommodate different loop strides for each type of core architecture.
\end{itemize}

The strategy is general and can be applied to a generic AMP,
consisting of any combination of fast/slow cores sharing the main memory, 
as well as to all other Level-3 BLAS operations.

To close this section, we emphasize that 
our work differs from~\cite{asymBLIS,OmpSsbigLITTLE} in that we address sophisticated DLA operations, 
with a rich hierarchy of task dependencies, by leveraging a conventional runtime scheduler in combination with
a data-parallel asymmetry-conscious implementation of the BLAS-3.

\section{Retargeting Existing Task Schedulers to Asymmetric Architectures}
\label{sec:runtime}

%
%
%
%

In this section, we initially perform an evaluation 
of the task-parallel Cholesky routine in Listings~\ref{lst:chol}--\ref{lst:chol_tasks},
executed on top of the conventional (i.e., default) scheduler in 
OmpSs linked to a sequential instance of BLIS,  on the target Exynos 5422 SoC.  The outcome from this study 
motivates the development effort and experiments presented in the remainder of the paper.

\subsection{Evaluation of conventional runtimes on AMPs}

\begin{figure}
\centering
\includegraphics[width=0.70\textwidth]{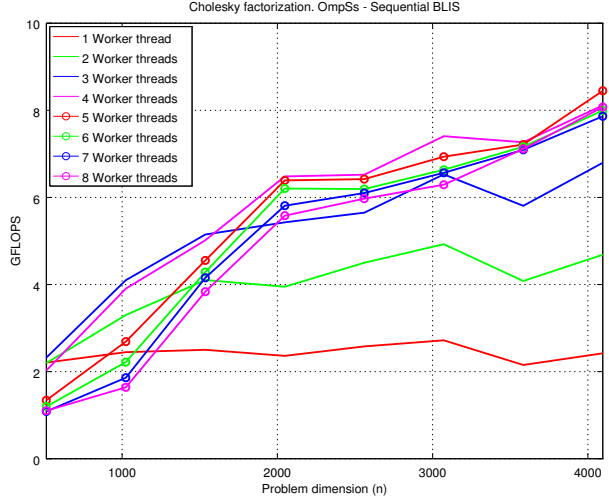}
\caption{Performance of the Cholesky factorization using the conventional OmpSs runtime and a sequential implementation
                 of BLIS on the Exynos 5422 SoC.}
\label{fig:ompss_blis_oversubscription}
\end{figure}

Figure~\ref{fig:ompss_blis_oversubscription} reports the performance,
in terms of GFLOPS (billions of flops per second), attained with the conventional OmpSs runtime,
when the number of worker threads varies from~1 to~8, and the mapping of worker threads to cores is delegated to the OS. 
We evaluated a range of block sizes
({\tt b} in Listing~\ref{lst:chol}), but for simplicity we report only the results obtained with the value {\tt b} that optimized
the GFLOPS rate for each problem dimension.
All the experiments hereafter employed {\sc ieee} double precision. Furthermore,
we  ensured that the cores operate at the highest possible frequency by setting the appropriate {\em cpufreq} governor.
The conventional runtime of OmpSs corresponds to release 15.06 of the Nanos++ runtime task scheduler.
For this experiment, it is lined with the ``sequential'' implementation of BLIS in release 0.1.5.
(For the experiments with the multi-threaded/asymmetric version of BLIS in the later sections, 
we will use specialized versions of the codes in~\cite{asymBLIS} for slow+fast VCs.)

The results in the
Figure reveal the increase in performance as the number of worker threads is raised from 1 to 4, which the OS maps
to the (big) Cortex-A15 cores. However, when the number of threads exceeds the amount of fast cores, the OS starts binding the threads
to the slower Cortex-A7 cores, and the improvement rate is drastically reduced, 
in some cases even showing a performance drop. This
is due to load imbalance, as tasks of uniform granularity, possibly laying in the critical path,
are assigned to slow cores. 

Stimulated by this first experiment, we recognize that an obvious solution to this problem consists in 
adapting the runtime task scheduler (more specifically, the scheduling policy) to exploit the 
SoC asymmetry~\cite{OmpSsbigLITTLE}.
Nevertheless, we part ways with this
solution, exploring an architecture-aware alternative
that leverages a(ny) conventional runtime task scheduler combined with an underlying asymmetric library. 
We discuss this option in more detail in the following section.

\subsection{Combining conventional runtimes with asymmetric libraries}

\subsubsection{General view}
Our proposal operates under the GTS model but is inspired in CPUM. Concretely,
our task scheduler regards
the computational resources as four {\em truly} symmetric VCs, each composed of a fast and a slow core. For this purpose, 
unlike CPUM, {\em both} physical cores within each VC remain active and collaborate to execute a given task.
Furthermore, our approach exploits concurrency at two levels: {\em task-level parallelism} is extracted by the runtime in order
to schedule tasks to the four symmetric VCs;
and each task/kernel is internally divided to 
expose {\em data-level parallelism}, distributing its workload between the two asymmetric physical cores within the VC
in charge of its execution.

Our solution thus only requires a conventional 
(and thus asymmetry-agnostic) runtime task scheduler, e.g. the conventional version of OmpSs, where instead of spawning one worker thread
per core in the system, we adhere to the CPUM model, creating only one worker thread per VC.
Internally, whenever a ready task is selected to be executed by a worker thread, 
the corresponding routine from BLIS internally spawns two threads, 
binds them to the appropriate pair of Cortex-A15+Cortex-A7 cores, 
and asymmetrically divides the work between the fast and the slow physical cores in the VC.
Following this idea, the architecture exposed to the runtime is {\em symmetric}, 
and the kernels in the BLIS library configure a ``black box'' that abstracts the 
architecture asymmetry from the runtime scheduler. 

In summary, in a conventional setup, the core is the basic computational resource for the task scheduler, 
and the ``sequential'' tasks are the minimum work unit to be assigned to these resources. Compared with this, in our approach the VC 
is the smallest (basic) computational resource from the point of view of the scheduler;
and tasks are further divided into smaller units, and executed in parallel by the physical cores inside the VCs.

\subsubsection{Comparison with other approaches}

Our approach features a number of advantages for the developer:
\begin{itemize}
\item The runtime is not aware of asymmetry, and thus a conventional task scheduler 
      will work transparently with no special modifications.
\item Any existing scheduling policy (e.g. cache-aware mapping or work stealing)
      in an asymmetry-agnostic runtime, or any enhancement technique, will directly impact the performance attained on an AMP.
\item Any improvement in the asymmetry-aware BLIS implementation will directly impact the performance on an AMP. 
      This applies to different ratios of big/LITTLE cores within a VC, operating frequency, or even to the introduction further levels
      of asymmetry (e.g. cores with a capacity between fast and slow).
\end{itemize}

Obviously, there is also a drawback in our proposal
as a tuned asymmetry-aware DLA library must exist in order to 
reuse conventional runtimes. 
In the scope of DLA, this drawback is easily tackled with BLIS. We recognize that, in more general domains, 
an ad-hoc implementation of the application's 
fundamental kernels becomes mandatory in order to fully exploit the underlying architecture.


\subsubsection{Requisites on the BLAS-3}

We finally note that certain requirements are imposed on a multi-threaded implementation of BLIS that operates under the 
CPUM mode. 
To illustrate this, consider the \gemm kernel and the high-level description of its implementation
in Listing~\ref{lst:gemm}. For our objective,
we still have to distribute the iteration space between the Cortex-A15 and the Cortex-A7 but, since there is only one resource of each type per VC,
there is no need to partition the loops internal to the macro-kernel. 
Furthermore, we note that the optimal strides for Loop~1 are in practice quite
large ({\tt nc} is in the order of a few thousands for ARM big.LITTLE cores), while the optimal values for Loop~3 are much more reduced
({\tt mc} is 32 for the Cortex-A7 and 156 for the Cortex-A15). Therefore, we target Loop~3 in our data-parallel implementation of BLIS for
VCs, which we can expect to easily yield a proper workload balancing.


\section{Experimental results}
\label{sec:results}

%

\subsection{Performance evaluation for asymmetric BLIS}

%

Let us start by reminding that, at execution time, OmpSs decomposes the routine for the Cholesky factorization into a collection of tasks 
that operate on sub-matrices (blocks) with a granularity 
dictated by the block size {\tt b}; see Listing~\ref{lst:chol}. 
These tasks typically perform invocations to a fundamental kernel of the BLAS-3, 
in our case provided by BLIS, or LAPACK; see Listing~\ref{lst:chol_tasks}.  

\newcommand{\bopt}{b^{\mbox{\rm \scriptsize opt}}\xspace}

The first step in our evaluation aims to provide a realistic estimation of the potential performance benefits
of our approach (if any) on the target Exynos 5422 SoC. 
A critical factor from this perspective is the range of block sizes, say $\bopt$,
that are optimal for the conventional OmpSs runtime. In particular, the efficiency 
of our hybrid task/data-parallel approach is strongly determined by the performance 
attained with the asymmetric BLIS implementation when compared against that of its sequential counterpart,
for problem dimensions {\tt n} that are in the order of $\bopt$.

Table~\ref{tab:optimal_bs_sym} reports the optimal block sizes $\bopt$ 
for the Cholesky factorization, with problems of increasing matrix dimension, using the conventional
OmpSs runtime linked with the sequential BLIS, and~1 to~4 worker threads.
Note that, except for smallest problems, the observed optimal block sizes
are between 192 and 448. These dimensions offer a 
fair compromise, exposing enough task-level parallelism
while delivering high ``sequential'' performance for the execution of each individual task
via the sequential implementation of BLIS.

%

\newcommand{\ra}[1]{\renewcommand{\arraystretch}{#1}}
\newcommand{\ca}[1]{\renewcommand{\tabcolsep}{#1}}

\ra{1.2}
\ca{2pt}

\begin{table}
	\centering
	\caption{Optimal block sizes for the Cholesky factorization using the conventional 
                 OmpSs runtime and a sequential implementation of BLIS on the Exynos 5422 SoC.}
	\label{tab:optimal_bs_sym}
{\scriptsize
\begin{tabular}{crrrrrrrrrrrrrrrr} 
\toprule
  & \phantom{a} & \multicolumn{14}{c}{Problem dimension ({\tt n})} \\ 
\cmidrule{3-17} 
  & \phantom{a} &     512 & 1,024 & 1,536 & 2,048 & 2,560 & 3,072 & 3,584 & 4,096 & 4,608 & 5,120 & 5,632 & 6,144 & 6,656 & 7,168 & 7,680 \\ \hline

{\sc 1 wt} & \phantom{a} &     192 & 384  & 320  & 448  & 448  & 448  & 384  & 320 & 320 & 448 & 448 & 448 & 448 & 384 & 448 \\ \hline
{\sc 2 wt} & \phantom{a} &     192 & 192  & 320  & 192  & 448  & 448  & 384  & 320 & 320 & 448 & 448 & 448 & 448 & 384 & 448 \\ \hline
{\sc 3 wt} & \phantom{a} &     128 & 192  & 320  & 192  & 384  & 448  & 320  & 320 & 320 & 448 & 448 & 448 & 448 & 384 & 448 \\ \hline
{\sc 4 wt} & \phantom{a} &     128 & 128  & 192  & 192  & 192  & 320  & 320  & 320 & 320 & 448 & 320 & 448 & 448 & 384 & 448 \\ \bottomrule
\end{tabular}
}
\end{table}

\begin{figure}[t]
\centering
\includegraphics[width=0.49\textwidth]{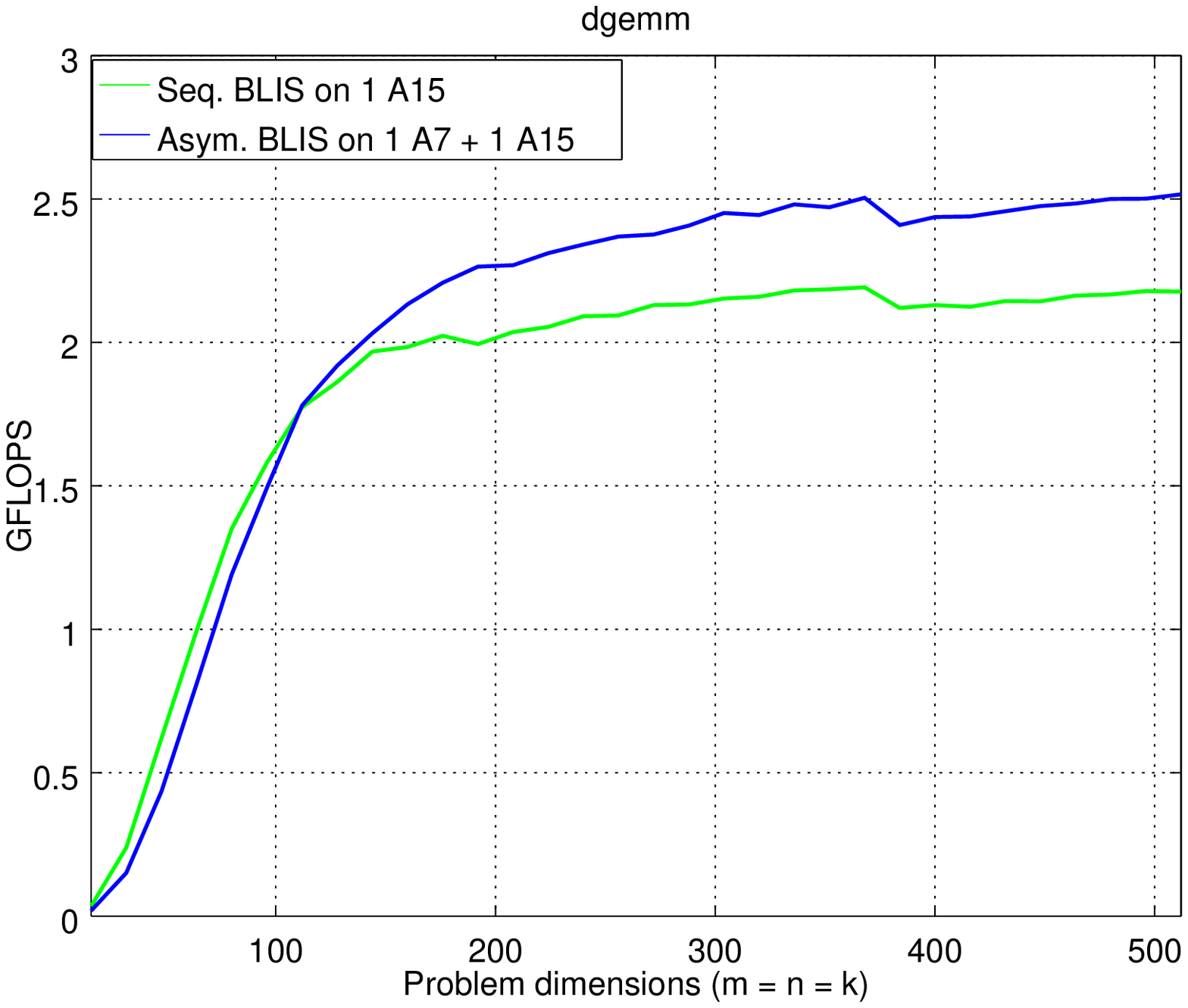}
\includegraphics[width=0.49\textwidth]{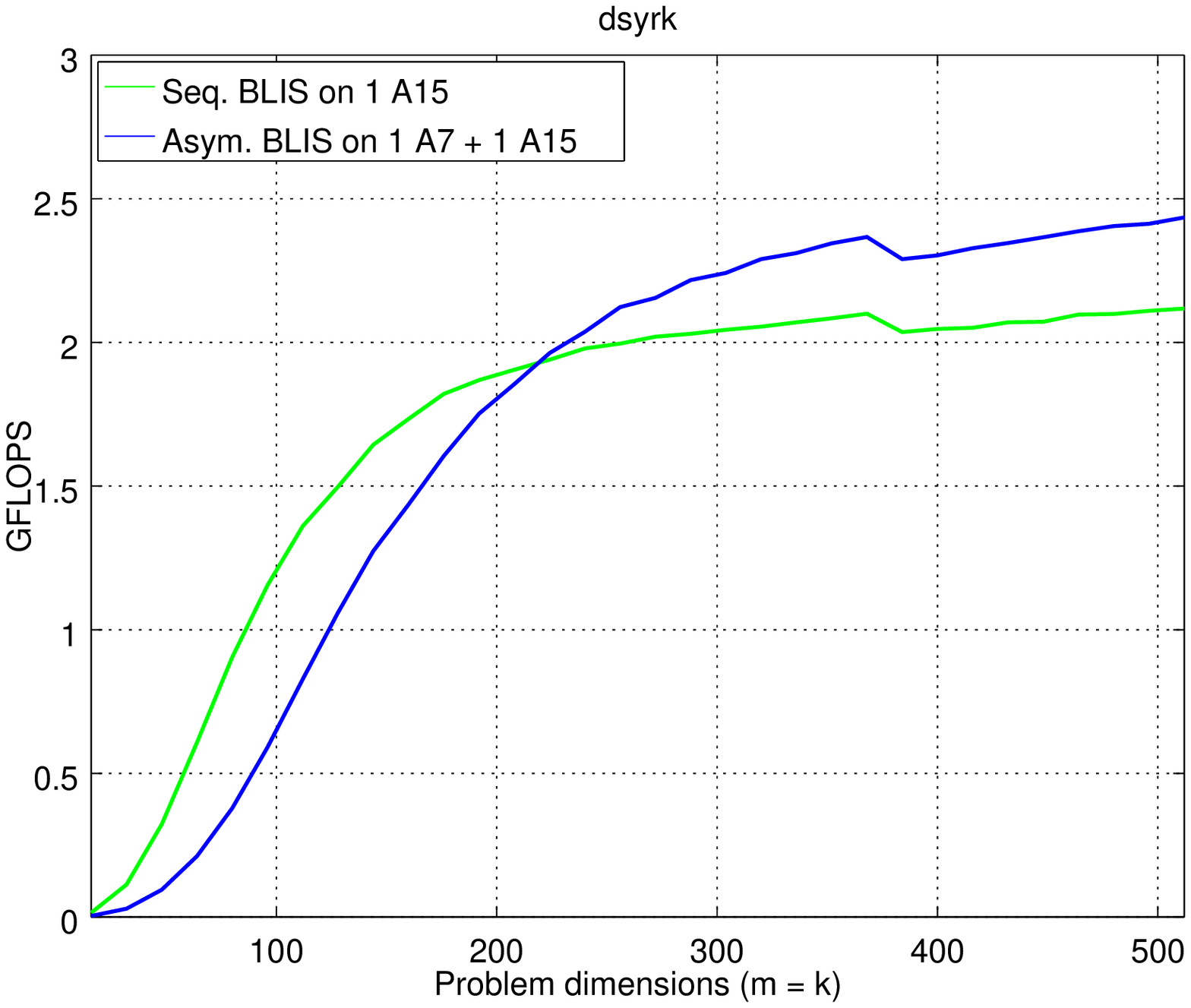}
\includegraphics[width=0.49\textwidth]{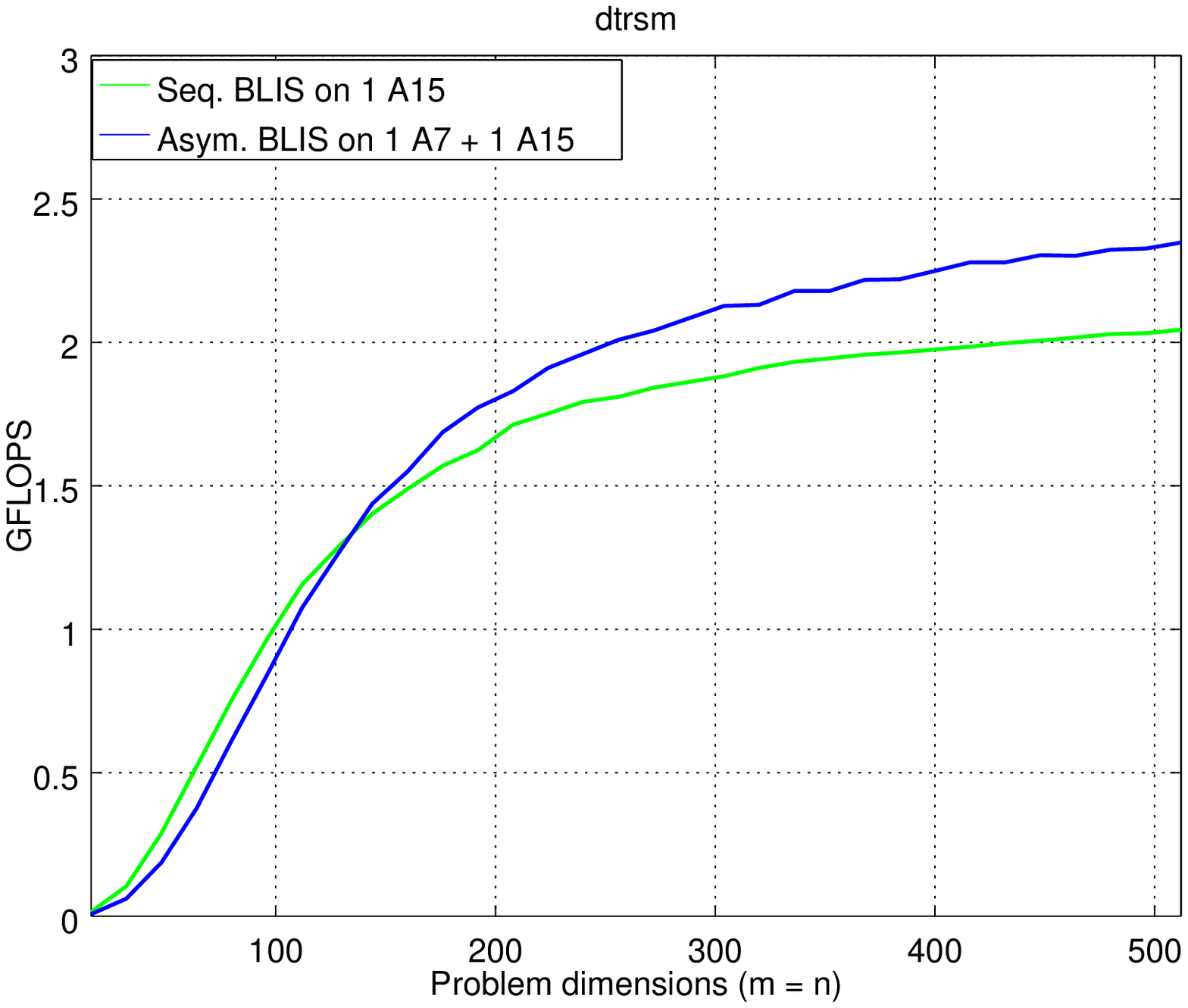}
\caption{Performance of the BLAS-3 kernels in the sequential and the multi-threaded/asymmetric implementations of
         BLIS, using respectively one Cortex-A15 core and one Cortex-A15 plus one Cortex-A7 core
         of the Exynos 5422 SoC.}
\label{fig:cross_blis}
\end{figure}

The key insight to take away from this experiments is that,
in order to extract good performance from a combination of the conventional OmpSs runtime task scheduler 
with a multi-threaded asymmetric version of BLIS, the kernels in this instance of the asymmetric library must outperform
their sequential counterparts, for matrix dimensions in the order of the block sizes in Table~\ref{tab:optimal_bs_sym}.  
Figure~\ref{fig:cross_blis} shows the performance attained
with the three BLAS-3 tasks involved in the Cholesky factorization (\gemm, \syrk and \trsm) for our range of dimensions of interest. 
There, the multi-threaded asymmetry-aware kernels run concurrently on one Cortex-A15 plus one Cortex-A7 core, while
the sequential kernels operate exclusively on a single Cortex-A15 core. 
In general, the three BLAS-3 routines exhibit a similar trend: the kernels from the sequential BLIS
outperform their asymmetric counterparts for small problems (up to 
approximately {\tt m},~{\tt n},~{\tt k}~=~128); but,
from that dimension, the use of the slow core starts paying off. The interesting aspect here is that
the cross-over threshold between both performance curves is in the range, (usually at an early point,) 
of $\bopt$; see Table~\ref{tab:optimal_bs_sym}. This implies 
that the asymmetric BLIS can potentially improve the performance of the overall computation. 
Moreover, the gap in performance grows with the problem size, 
stabilizing at problem sizes around {\tt m},~{\tt n},~{\tt k}~$\approx~400$. 
Given that this value is in the range of the optimal block size for the task-parallel Cholesky implementation, 
we can expect a performance increment in the order of 0.3--0.5 GFLOPS per added slow core, 
mimicking the behavior of the underlying BLIS.

\subsection{Integration of asymmetric BLIS in a conventional task scheduler}

\begin{figure}[t]
\centering
\includegraphics[width=0.49\textwidth]{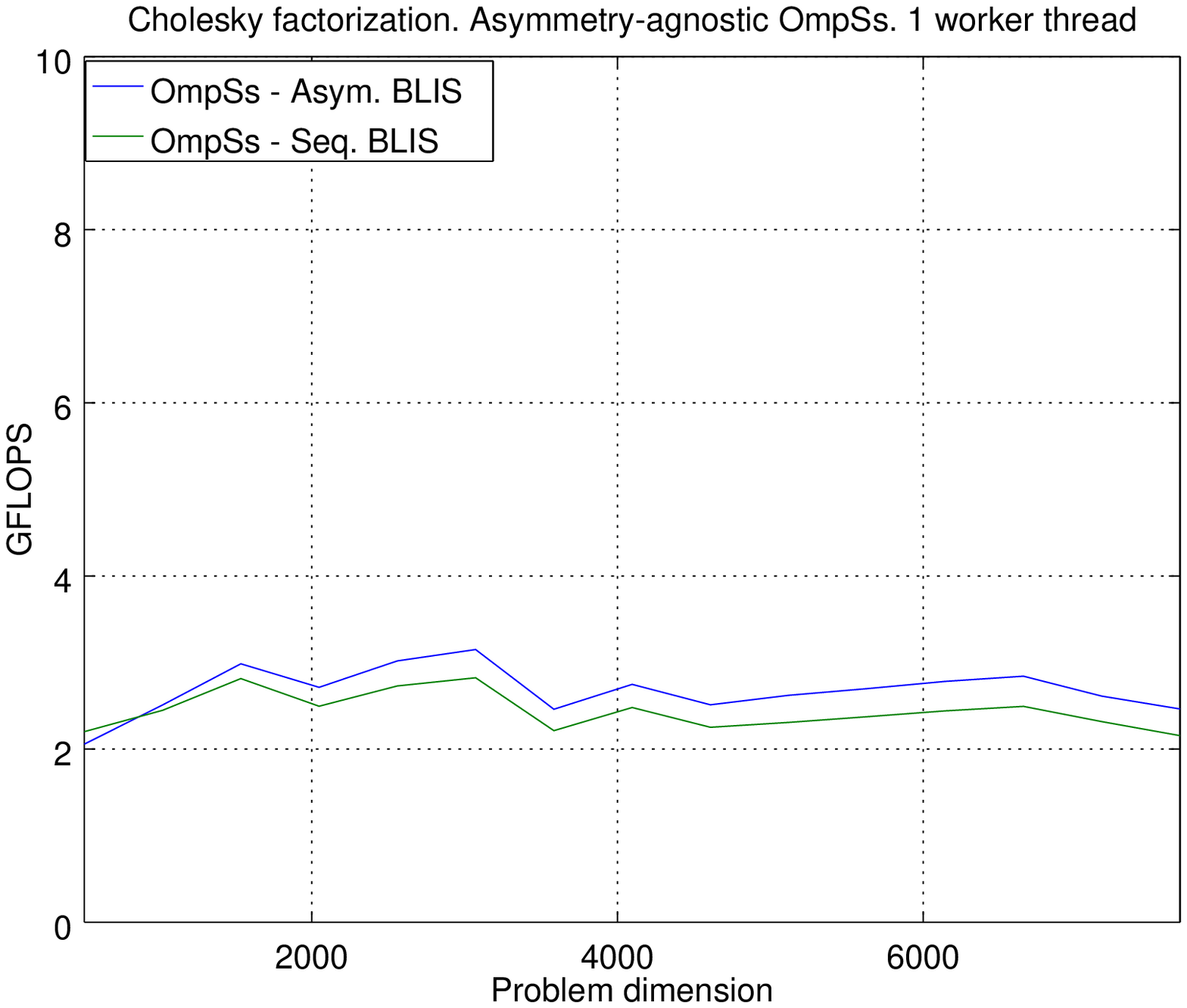}
\includegraphics[width=0.49\textwidth]{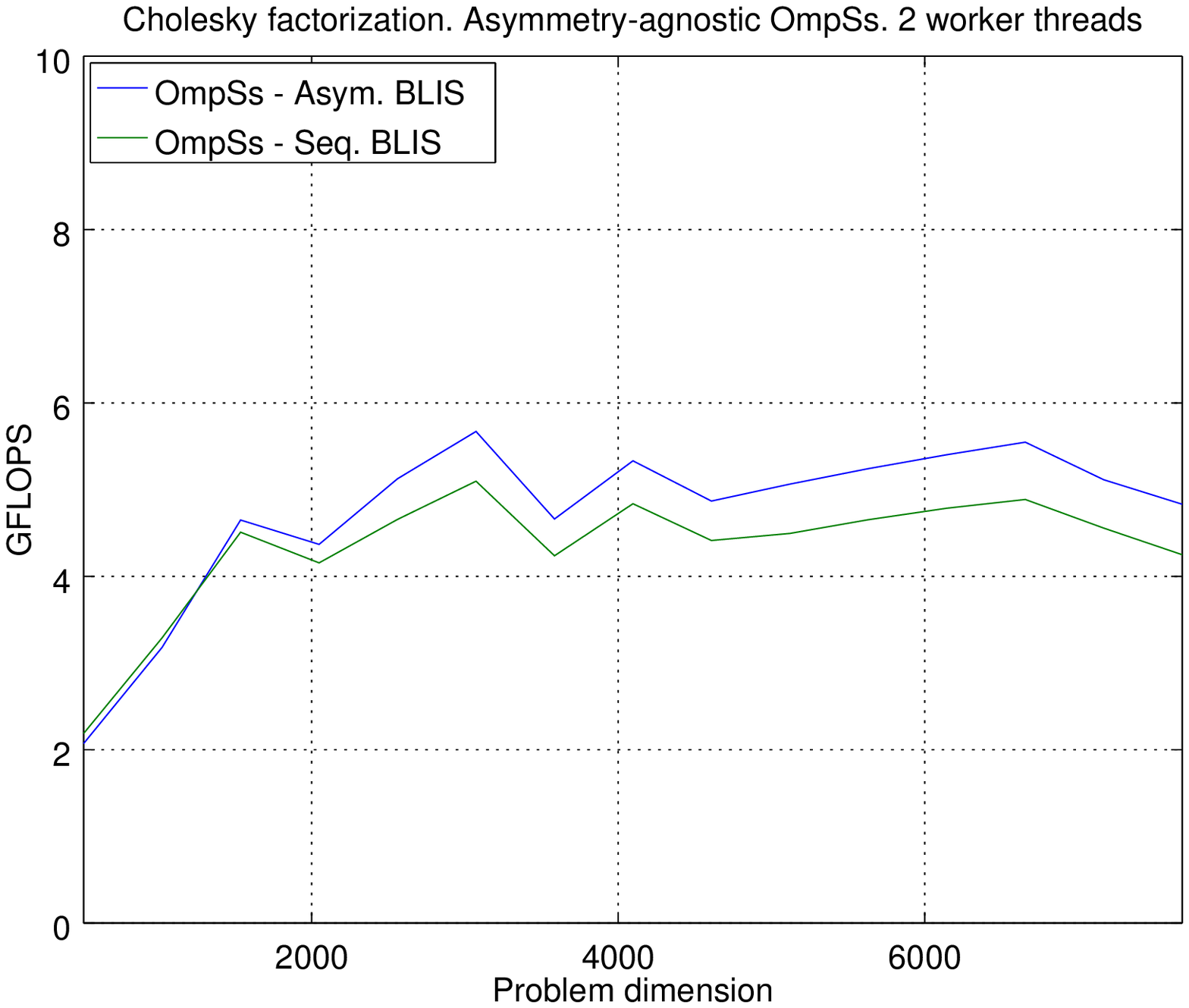}
\includegraphics[width=0.49\textwidth]{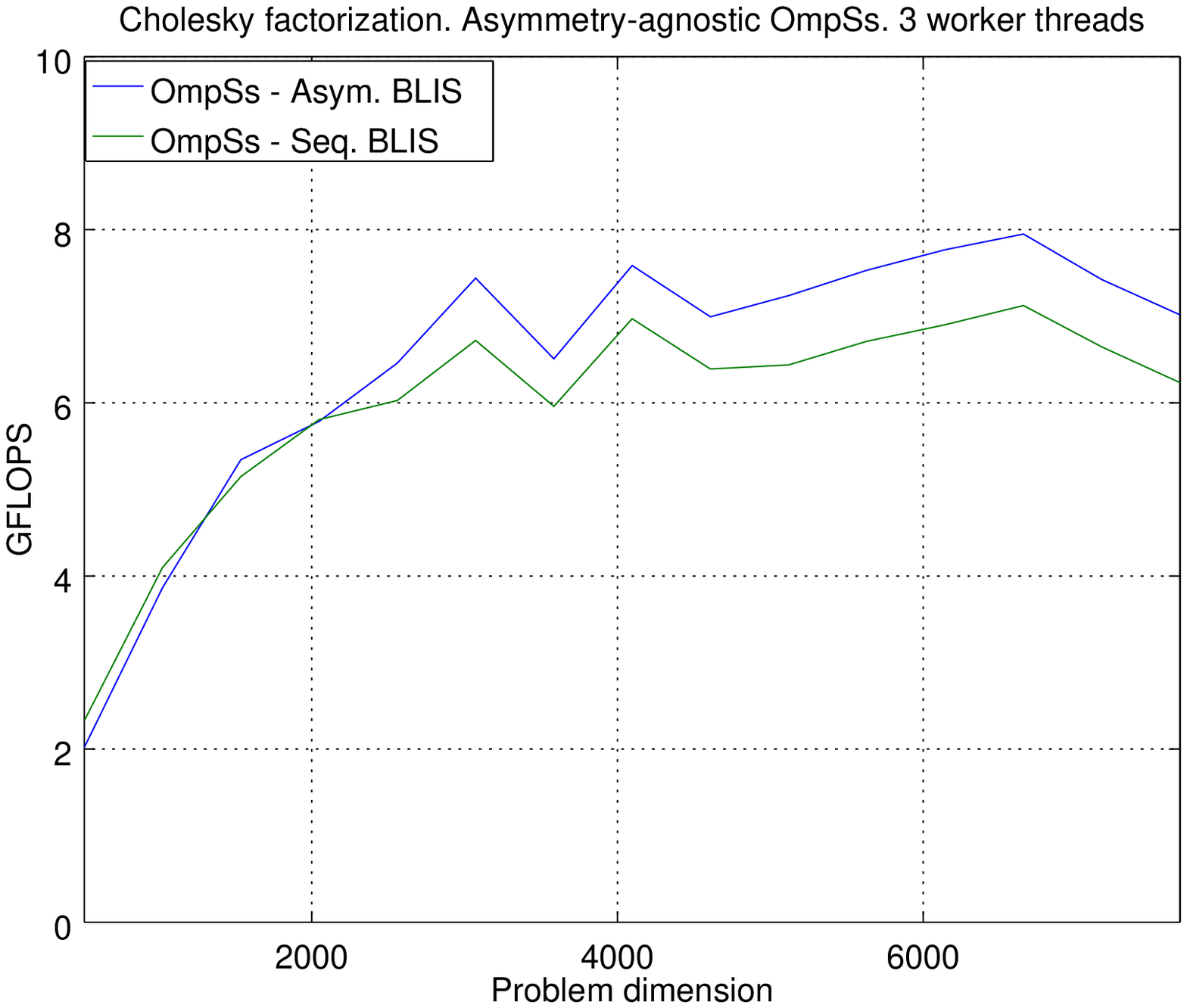}
\includegraphics[width=0.49\textwidth]{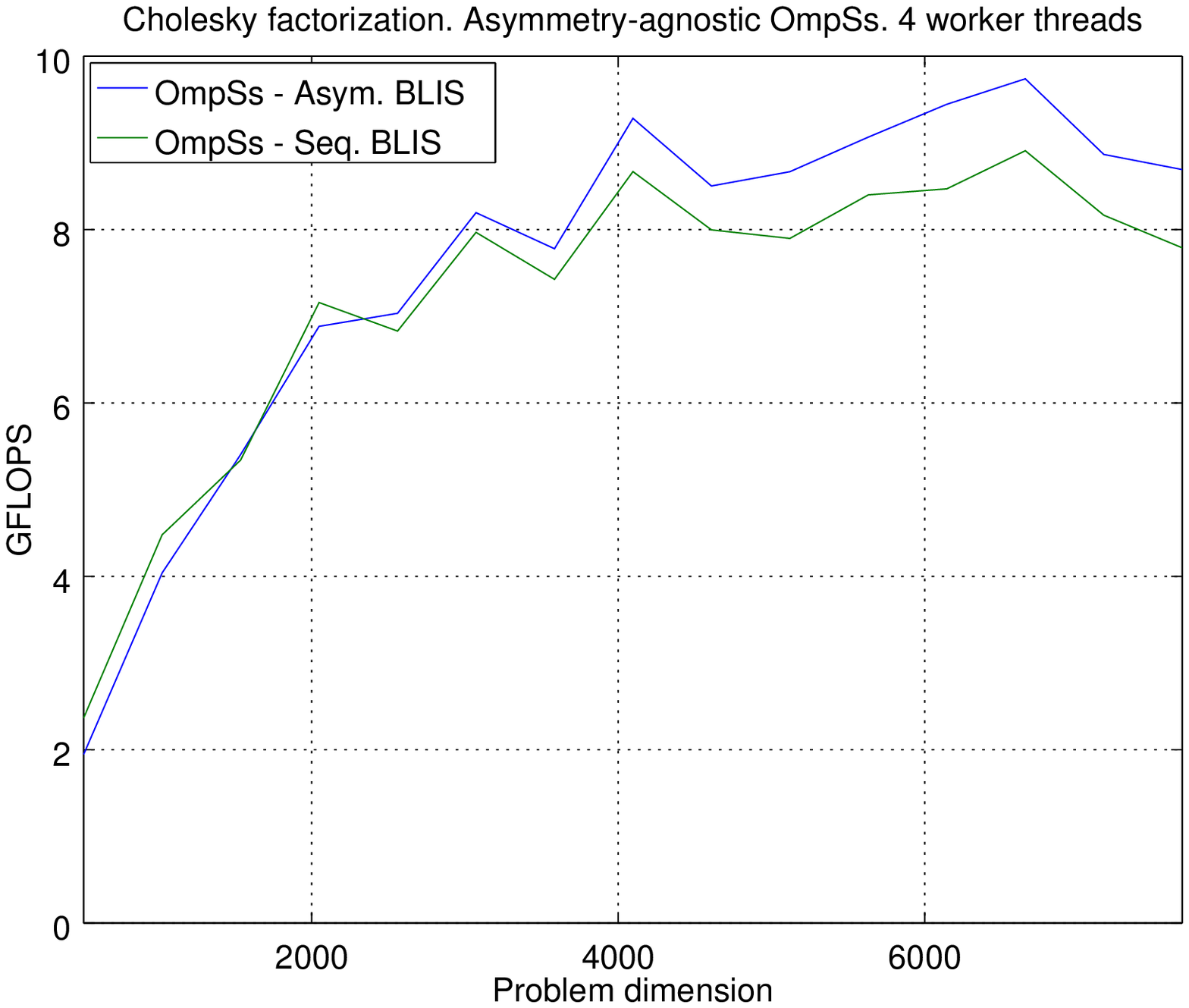}
\caption{Performance of the Cholesky factorization using the conventional OmpSs runtime linked with either
         the sequential or the multi-threaded/asymmetric implementations of
         BLIS 
         in the Exynos 5422 SoC.}
\label{fig:ompss_blis}
\end{figure}

In order to analyze the actual benefits of our proposal, we next evaluate the conventional 
OmpSs task scheduler linked with either the sequential implementation of BLIS or its multi-threaded asymmetry-aware version. 
Hereafter, the BLIS kernels from first configuration always run using one Cortex-A15 core while, in the second case,
they exploit one Cortex-A15 plus one Cortex-A7 core.
Figure~\ref{fig:ompss_blis} reports the results for both setups, using an increasing number 
of worker threads from~1 to~4. For simplicity, we only report the results obtained with the optimal block size. 
In all cases, the solution based on the multi-threaded asymmetric library outperforms the sequential implementation for 
relatively large matrices (usually for dimensions {\tt n}~$>$~2,048) while, for smaller problems, the GFLOPS rates are 
similar. The reason for this behavior can be derived from the optimal block sizes reported in  
Table~\ref{tab:optimal_bs_sym} and the performance of BLIS reported in Figure~\ref{fig:cross_blis}: for that range 
of problem dimensions, the optimal block size is significantly smaller, and both BLIS implementations attain similar
performance rates.

%
%

The quantitative difference in performance between both approaches is
reported in Tables~\ref{tab:improvement_absolute} and~\ref{tab:improvement_percore}. 
The first table illustrates the raw (i.e., absolute) gap, while the second
one shows the difference per Cortex-A7 core introduced in the experiment. 
Let us consider, for example, the problem size {\tt n}~=~6,144. 
In that case, the performance roughly improves by 0.975 GFLOPS when the 4~slow cores are added to help the base 4~Cortex-A15 cores. 
This translates into a performance raise of 0.243 GFLOPS per slow core, which is slightly under the 
improvement that could be expected from results experiments in the previous section. 
Note, however, that the performance per Cortex-A7 core is reduced from 0.340~GFLOPS, when adding just one
core, to 0.243~GFLOPS, when simultaneously using all four slow cores.

\newcommand{\fg}[1]{\textcolor{ForestGreen}{#1}} 
\newcommand{\br}[1]{\textcolor{BrickRed}{#1}} 

\begin{table}
	\centering
\caption{Absolute performance improvement (in GFLOPS) for the Cholesky factorization using
         the conventional OmpSs runtime linked with 
         the multi-threaded/asymmetric BLIS with respect to the same runtime linked with the sequential BLIS in
         the Exynos 5422 SoC.}
	 \label{tab:improvement_absolute}


\ra{1.2}
\ca{2pt}

{\scriptsize
\begin{tabular}{crrrrrrrrrrrrr} 
   	\toprule
                 & \phantom{a} & \multicolumn{12}{c}{Problem dimension ({\tt n})} \\ 
\cmidrule{3-14} 
     & \phantom{a} &       512      & 1,024        & 2,048          & 2,560        & 3,072        & 4,096        & 4,608        & 5,120        & 5,632        & 6,144        & 6,656         & 7,680 \\ \hline 
{\sc 1 wt} & \phantom{a} &    \br{-0.143} & \fg{0.061}  & \fg{0.218}    & \fg{0.289}  & \fg{0.326}  & \fg{0.267}  & \fg{0.259}  & \fg{0.313}  & \fg{0.324}  & \fg{0.340}  & \fg{0.348}   & \fg{0.300} \\ \cline{3-14}
{\sc 2 wt} & \phantom{a} &    \br{-0.116} & \br{-0.109} & \fg{0.213}    & \fg{0.469}  & \fg{0.573}  & \fg{0.495}  & \fg{0.454}  & \fg{0.568}  & \fg{0.588}  & \fg{0.617}  & \fg{0.660}   & \fg{0.582} \\ \cline{3-14}
{\sc 3 wt} & \phantom{a} &    \br{-0.308} & \br{-0.233} & \br{-0.020}   & \fg{0.432}  & \fg{0.720}  & \fg{0.614}  & \fg{0.603}  & \fg{0.800}  & \fg{0.820}  & \fg{0.866}  & \fg{0.825}   & \fg{0.780} \\ \cline{3-14}
{\sc 4 wt} & \phantom{a} &    \br{-0.421} & \br{-0.440} & \br{-0.274}   & \fg{0.204}  & \fg{0.227}  & \fg{0.614}  & \fg{0.506}  & \fg{0.769}  & \fg{0.666}  & \fg{0.975}  & \fg{0.829}   & \fg{0.902} \\ \bottomrule
\end{tabular}
}
\end{table}

%
%

\begin{table}
\centering
\caption{Performance improvement per slow core (in GFLOPS) for the Cholesky factorization using
         the conventional OmpSs runtime linked with 
         the multi-threaded/asymmetric BLIS with respect to the same runtime linked with the sequential BLIS in
         the Exynos 5422 SoC.}
\label{tab:improvement_percore}

\ra{1.2}
\ca{2pt}
{\scriptsize
\begin{tabular}{crrrrrrrrrrrrr} 
   	\toprule
                 & \phantom{a} & \multicolumn{12}{c}{Problem dimension ({\tt n})} \\ 
\cmidrule{3-14} 
     & \phantom{a} &       512      & 1,024        & 2,048          & 2,560        & 3,072        & 4,096        & 4,608        & 5,120        & 5,632        & 6,144        & 6,656         & 7,680 \\ \hline 
	 {\sc 1 wt}    & \phantom{a} &    \br{-0.143} & \fg{0.061}  & \fg{0.218}  & \fg{0.289} & \fg{0.326}  & \fg{0.267} & \fg{0.259} & \fg{0.313} & \fg{0.324} & \fg{0.340} & \fg{0.348} & \fg{0.300}    \\ \cline{3-14}
	 {\sc 2 wt}    & \phantom{a} &    \br{-0.058} & \br{-0.054} & \fg{0.106}  & \fg{0.234} & \fg{0.286}  & \fg{0.247} & \fg{0.227} & \fg{0.284} & \fg{0.294} & \fg{0.308} & \fg{0.330} & \fg{0.291}    \\ \cline{3-14}
	 {\sc 3 wt}    & \phantom{a} &    \br{-0.102} & \br{-0.077} & \br{-0.006} & \fg{0.144} & \fg{0.240}  & \fg{0.204} & \fg{0.201} & \fg{0.266} & \fg{0.273} & \fg{0.288} & \fg{0.275} & \fg{0.261}    \\ \cline{3-14}
	 {\sc 4 wt}    & \phantom{a} &    \br{-0.105} & \br{-0.110} & \br{-0.068} & \fg{0.051} & \fg{0.056}  & \fg{0.153} & \fg{0.126} & \fg{0.192} & \fg{0.166} & \fg{0.243} & \fg{0.207} & \fg{0.225}    \\ \bottomrule
\end{tabular}
}

\end{table}

\subsection{Performance comparison versus asymmetry-aware task scheduler}
\label{sec:comparative}

Our last round of experiments aims to assess the performance advantages of different task-parallel executions
of the Cholesky factorization via OmpSs. Concretely, we consider 
(1) the conventional task scheduler linked with the sequential BLIS (``OmpSs - Seq. BLIS''); 
(2) the conventional task scheduler linked with our multi-threaded asymmetric BLIS that views the SoC as four symmetric {\em virtual cores}
(``OmpSs - Asym. BLIS''); and 
(3) the criticality-aware task scheduler in Botlev-OmpSs linked with the sequential BLIS
(``Botlev-OmpS - Seq. BLIS''). 
In the executions, we use all four Cortex-A15 cores and 
evaluate the impact of adding an increasing number of Cortex-A7 cores, from~1 to~4, for Botlev-OmpSs. 

\begin{figure}
\centering
\includegraphics[width=0.70\textwidth]{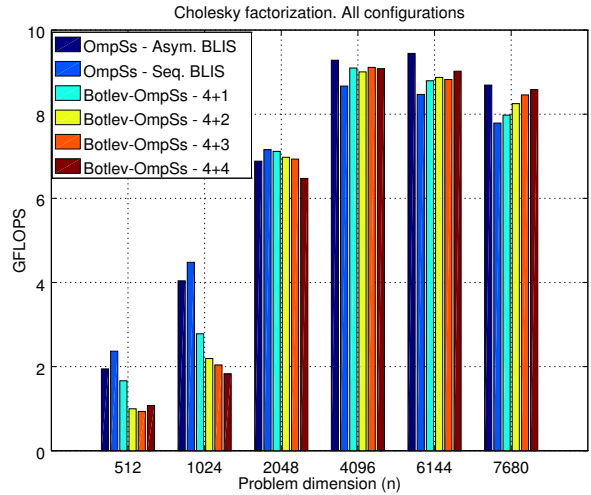}
\caption{Performance (in GFLOPS) for the Cholesky factorization using
         the conventional OmpSs runtime linked with 
         either the sequential BLIS or the multi-threaded/asymmetric BLIS, and the {\em ad-hoc} asymmetry-aware version of the
         OmpSs runtime (Botlev-OmpSs) linked with the sequential BLIS in 
         the Exynos 5422 SoC. The labels of the form ``4+x'' refer to an execution with 4 Cortex-A15 cores and x Cortex-A7 cores.}
\label{fig:comparative}
\end{figure}

Figure~\ref{fig:comparative} shows the performance attained by the aforementioned alternatives on the Exynos 5422 SoC.
The results can be divided into groups along three problem dimensions:
\begin{itemize}
 \item For small matrices ({\tt n}~=~512, 1,024), the conventional runtime using exclusively the four big cores (that is,
linked with a sequential BLIS library for task execution) attains the best results in terms of performance. This was
expected and was already observed in Figure~\ref{fig:ompss_blis}; the main reason is the small optimal block size,
enforced by the reduced problem size, that is necessary in order to expose enough task-level parallelism. This invalidates the
use of our asymmetric BLIS implementation due to the low performance for very small matrices;
see Figure~\ref{fig:cross_blis}. We note that the ad-hoc Botlev-OmpSs does not attain
remarkable performances either for this dimension range, regardless the amount of Cortex-A7 cores used.

 \item For medium-sized matrices ({\tt n}~=~2,048, 4,096), the gap in performance between the different approaches is reduced. 
The variant that relies on the asymmetric BLIS implementation commences to outperform the alternative implementations for
{\tt n}=4,096 by a short margin. For this problem range, Botlev-OmpSs is competitive, and also outperforms the conventional setup.

 \item For large matrices ({\tt n}~=~6,144, 7,680) this trend is consolidated, and both asymmetry-aware approaches deliver
remarkable performance gains with respect to the conventional setup. Comparing both asymmetry-aware solutions, our
mechanism attains better performance rate, even when considering the usage of all available cores for the Botlev-OmpSs
runtime version.

\end{itemize}

To summarize, our proposal to exploit asymmetry improves portability and programmability by avoiding
a revamp of the runtime task scheduler for AMPs. In addition, our approach renders performance
gains which are, for all problems cases, comparable with those of ad-hoc asymmetry-conscious schedulers; 
for medium to large matrices, it clearly outperforms the efficiency attained with a conventional 
asymmetry-oblivious scheduler.

\subsection{Extended performance analysis}

We next provide further details on the performance behavior of each one of the aforementioned runtime configurations.
The execution traces in this section have all been extracted with the {\tt Extrae} instrumentation
tool and analyzed with the visualization package {\tt Paraver}~\cite{Paraver}. 
The results correspond to the Cholesky factorization of a single problem with matrix dimension {\tt n}~=~6,144 and 
block size {\tt b}~=~448.

\subsubsection{General task execution overview.}

\begin{figure}
\centering
 \subfigure[OmpSs - Sequential BLIS (8 worker threads)]{
   \includegraphics[width=\textwidth]{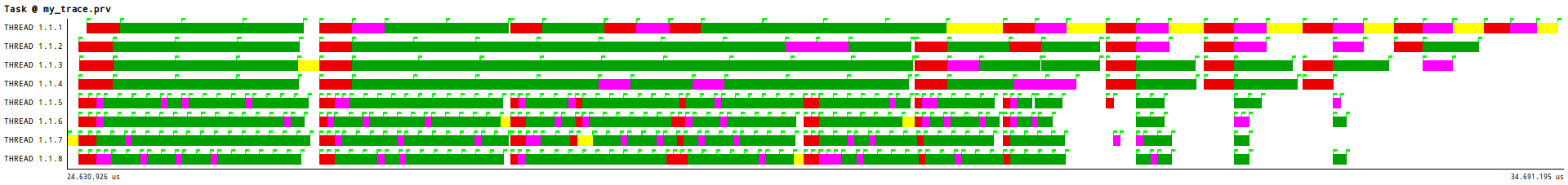}
}
 \subfigure[OmpSs - Sequential BLIS (4 worker threads)]{
   \includegraphics[width=\textwidth]{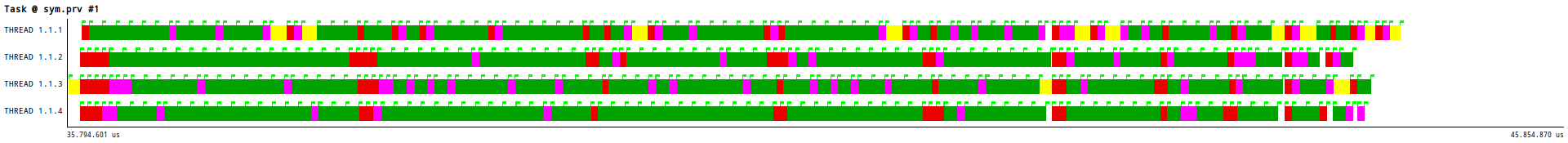}
}
 \subfigure[OmpSs - Asymmetric BLIS (4 worker threads)]{
   \includegraphics[width=\textwidth]{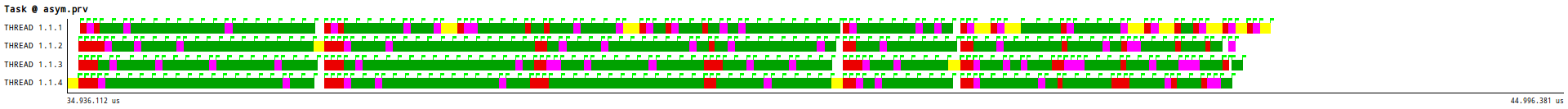}
}
 \subfigure[Botlev-OmpSs - Sequential BLIS (8 worker threads, 4+4)]{
   \includegraphics[width=\textwidth]{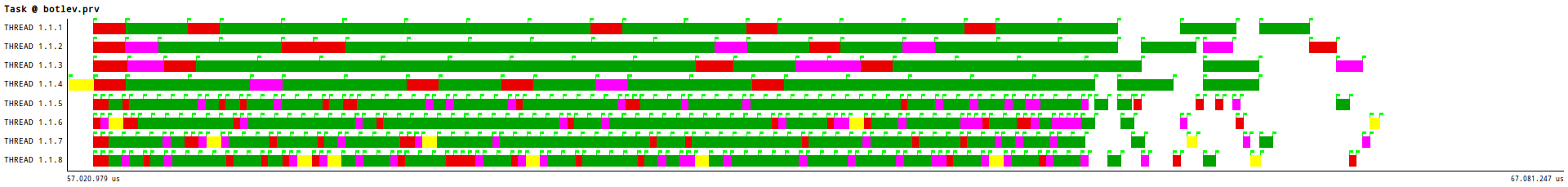}
}
\caption{Execution traces of the three runtime configurations for the Cholesky factorization ({\tt n}~=~6,144, 
{\tt b}~=~448). 
The timeline in each row collects the tasks executed by a single worker thread. 
Tasks are colored following the convention in Figure~\ref{fig:dag};
phases colored in white between tasks represent idle times.
The green flags mark task initialization/completion. 
}
\label{fig:traces_tasks}
\end{figure}

Figure~\ref{fig:traces_tasks} reports complete execution traces for each runtime configuration. 
At a glance, a number of coarse remarks can be
extracted from the trace:
\begin{itemize}
\item From the perspective of total execution time (i.e., {\em time-to-solution}), the conventional OmpSs runtime combined with 
      the asymmetric BLIS implementation attains the best results, followed by the Botlev-OmpSs runtime configuration. It is worth pointing out that
      an asymmetry-oblivious runtime which spawns 8~worker threads, with no further considerations, yields the worst performance by far. In this case, the
      load imbalance and long idle periods, especially as the amount of concurrency decreases in the final part of the trace, entail
      a huge performance penalty. 
\item The flag marks indicating task initialization/completion reveal that 
      the asymmetric BLIS implementation (which employs the combined resources from a VC) requires less time per task than 
      the two alternatives based on a sequential BLIS. An effect to note specifically in the Botlev-OmpSs configuration is the
      difference in performance between tasks of the same type, when
      executed by a big core (worker threads~5 to~8) or a LITTLE one (worker threads~1 to~4).
\item The Botlev-OmpSs task scheduler embeds a (complex) scheduling strategy that includes priorities,
      advancing the execution of tasks in the critical path and, whenever possible, assigning them to fast cores (see, for 
      example, the tasks for the factorization of diagonal blocks, 
      colored in yellow). This yields an execution timeline that is more compact during the first stages of the parallel execution, 
      at the cost of longer idle times when the degree of concurrency decreases (last iterations of the factorization). 
      Although possible, a priority-aware technique has not been applied in our experiments with the conventional OmpSs setups
      and remains part of future work.
\end{itemize}

We next provide a quantitative analysis on the task duration and a more detailed study of 
the scheduling strategy integrated in each configuration.

\subsubsection{Task duration.}
 
Table~\ref{tab:2dp_tasks} reports the average execution time per type of task for each worker thread.
These results show that the execution time per individual type of task is considerably shorter
for our multithreaded/asymmetric BLIS implementation than for the alternatives based on a sequential
version of BLIS. The only exception is the factorization of the diagonal block ({\tt dpotrf}) as this is
an LAPACK-level routine, and therefore it is not available in BLIS. Inspecting the task execution time of
the Botlev-OmpSs configuration, we observe a remarkable difference depending
on the type of core tasks are mapped to. For example, the average execution times for {\tt dgemm}
range from more than 400~ms on a LITTLE core, to roughly 90~ms on a big core. This
behavior is reproduced for all types of tasks.

\begin{table}
\centering
\caption{Average time (in ms) per task and worker thread in the Cholesky factorization  ({\tt n}~=~6,144, 
{\tt b}~=~448), for the three runtime configurations.}
\label{tab:2dp_tasks}

\ra{1.2}
\ca{2pt}

\renewcommand{\fg}[1]{{#1}} 
\renewcommand{\br}[1]{{#1}} 

{\scriptsize
\begin{tabular}{crrrrrrrrrrrrrrr} 
   	\toprule
                 & \phantom{a} & \multicolumn{4}{c}{OmpSs - Seq. BLIS} & \phantom{ab} & \multicolumn{4}{c}{OmpSs - Asym. BLIS} & \phantom{ab} & \multicolumn{4}{c}{Botlev-OmpSs - Seq. BLIS} \\ 
                 & \phantom{a} & \multicolumn{4}{c}{(4 worker threads)} & \phantom{ab} & \multicolumn{4}{c}{(4 worker threads)} & \phantom{ab} & \multicolumn{4}{c}{(8 worker threads, 4+4)} \\ 
                                          \cmidrule{3-6}                                         \cmidrule{8-11}                                      \cmidrule{13-16}
                       & \phantom{a} &    {\tt dgemm} & {\tt dtrsm}& {\tt dsyrk}& {\tt dpotrf}  & \phantom{ab} & {\tt dgemm}  & {\tt dtrsm} & {\tt dsyrk} & {\tt dpotrf}& \phantom{ab} & {\tt dgemm} & {\tt dtrsm} & {\tt dsyrk} & {\tt dpotrf}         \\ \hline 
	 {\sc wt 0}    & \phantom{a} &    \br{89.62} & \fg{48.12} & \fg{47.14} & \fg{101.77}    & \phantom{ab} & \fg{79.82}  & \fg{42.77}  & \fg{44.42}  & \fg{105.77}  & \phantom{ab} & \fg{406.25} & \fg{216.70} & \fg{--} & \fg{--}    \\ \cline{3-16}
	 {\sc wt 1}    & \phantom{a} &    \br{88.96} & \br{48.10} & \fg{47.14} & \fg{--}      & \phantom{ab} & \fg{78.65}  & \fg{42.97}  & \fg{44.56}  & \fg{76.35}  & \phantom{ab} & \fg{408.90} & \fg{207.41} & \fg{212.55} & \fg{--}    \\ \cline{3-16}
	 {\sc wt 2}    & \phantom{a} &    \br{89.02} & \br{48.36} & \br{47.18} & \fg{87.22}    & \phantom{ab} & \fg{79.14}  & \fg{43.14}  & \fg{44.60}  & \fg{85.98}  & \phantom{ab} & \fg{415.31} & \fg{230.07} & \fg{212.56} & \fg{--}    \\ \cline{3-16}
	 {\sc wt 3}    & \phantom{a} &    \br{90.11} & \br{48.51} & \br{47.42} & \fg{--}      & \phantom{ab} & \fg{79.28}  & \fg{43.10}  & \fg{44.59}  & \fg{67.73}  & \phantom{ab} & \fg{410.84} & \fg{216.95} & \fg{216.82} & \fg{137.65}    \\ \cline{3-16}
	 {\sc wt 4}    & \phantom{a} &    \br{--} & \br{--} & \br{--} & \fg{--} & \phantom{ab} & \fg{--}  & \fg{--} & \fg{--} & \fg{--} & \phantom{ab} & \fg{90.97} & \fg{48.97} & \fg{48.36} & \fg{--}    \\ \cline{3-16}
	 {\sc wt 5}    & \phantom{a} &    \br{--} & \br{--} & \br{--} & \fg{--} & \phantom{ab} & \fg{--}  & \fg{--} & \fg{--} & \fg{--} & \phantom{ab} & \fg{90.61} & \fg{48.86} & \fg{48.16} & \fg{90.78}    \\ \cline{3-16}
	 {\sc wt 6}    & \phantom{a} &    \br{--} & \br{--} & \br{--} & \fg{--} & \phantom{ab} & \fg{--}  & \fg{--} & \fg{--} & \fg{--} & \phantom{ab} & \fg{91.28} & \fg{49.43} & \fg{47.97} & \fg{89.58}    \\ \cline{3-16}
	 {\sc wt 7}    & \phantom{a} &    \br{--} & \br{--} & \br{--} & \fg{--} & \phantom{ab} & \fg{--}  & \fg{--} & \fg{--} & \fg{--} & \phantom{ab} & \fg{91.60} & \fg{49.49} & \fg{48.62} & \fg{95.43}    \\ \bottomrule
	 {\sc Avg.}     & \phantom{a} &    \br{89.43} & \br{48.27} & \br{47.22} & \fg{94.49} & \phantom{ab} & \fg{79.22}   & \fg{42.99} & \fg{44.54} & \fg{83.96} & \phantom{ab} & \fg{250.72} & \fg{133.49} & \fg{119.29} & \fg{103.36}    \\ \bottomrule
\end{tabular}
}

\end{table}

\subsubsection{Task scheduling policies and idle times.}

\begin{figure}[t]
\centering
 \subfigure[OmpSs - Sequential BLIS (8 worker threads)]{
   \includegraphics[width=\textwidth]{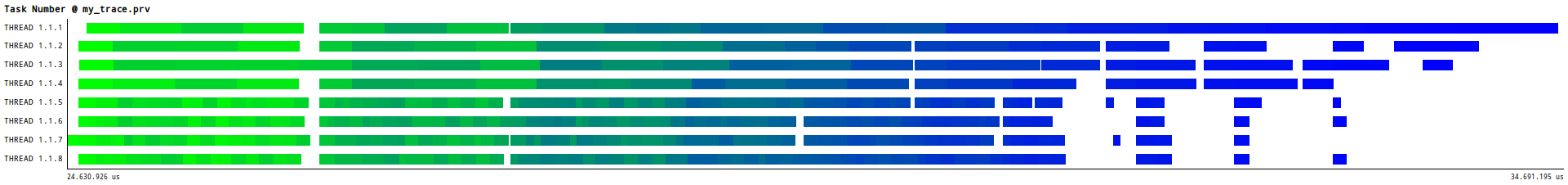}
}
 \subfigure[OmpSs - Sequential BLIS (4 worker threads)]{
   \includegraphics[width=\textwidth]{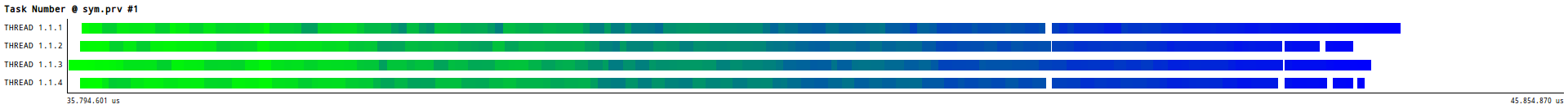}
}
 \subfigure[OmpSs - Asymmetric BLIS (4 worker threads)]{
   \includegraphics[width=\textwidth]{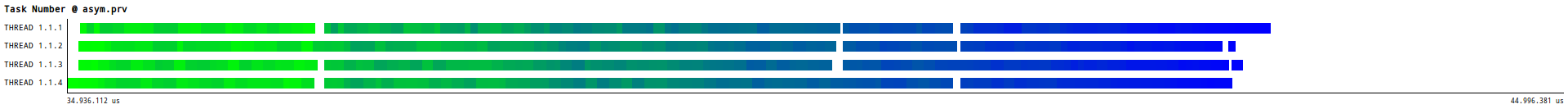}
}
 \subfigure[Botlev-OmpSs - Sequential BLIS (8 worker threads, 4+4)]{
   \includegraphics[width=\textwidth]{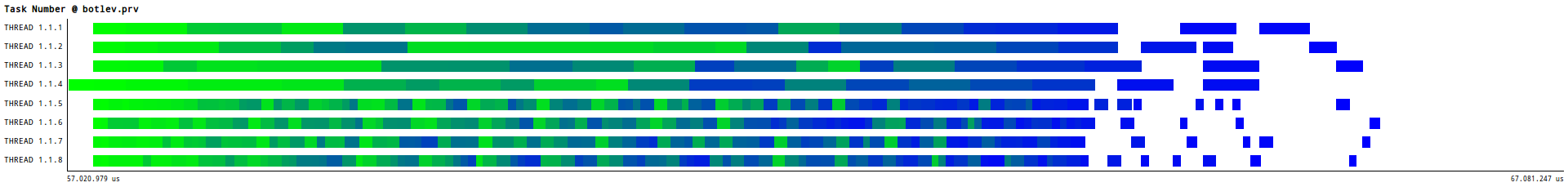}
}
\caption{Task execution order of the three studied runtime configurations for the Cholesky factorization 
({\tt n}~=~6,144, {\tt b}~=~448). In the trace, tasks are ordered
according to their appearance in the sequential code, 
and depicted using a color gradient, with light green indicating early tasks, and dark blue for the late
tasks. 
}
\label{fig:traces_task_number}
\end{figure}

Figure~\ref{fig:traces_task_number} illustrates the task execution order determined by the Nanos++ task scheduler. 
Here tasks are depicted using a color gradient, attending to the order in which they are encountered in the sequential code, 
from the earliest to the latest.

At runtime, the task scheduler in Botlev-OmpSs issues tasks to execution out-of-order depending on their criticality. 
The main idea behind this scheduling policy is to track the criticality of each task and, when possible, 
advance the execution of critical tasks assigning them to the fast Cortex-A15 cores. 
Conformally with this strategy, an out-of-order execution reveals itself more frequently in the timelines for the big cores 
than in those for the LITTLE cores. 
With the conventional runtime, the out-of-order execution is only dictated 
by the order in which data dependencies for tasks are satisfied.

From the execution traces, we can observe that the Botlev-OmpSs alternative suffers a remarkable
performance penalty due to the existence of idle periods in the final part of the factorization, 
when the concurrency in the factorization is greatly diminished. 
This problem is not present in the conventional scheduling policies.
In the first stages of the factorization, however, the use of a priority-aware policy for the Botlev-OmpSs scheduler
effectively reduces idle times. 
Table~\ref{tab:th_state} reports the percentage of time each worker
thread is in {\tt running} or {\tt idle} state. In general, the relative amount of time spent in idle state is much 
higher for Botlev-OmpSs than for the conventional implementations (17\% vs. 5\%, respectively). 
Note also the remarkable difference in the percentage of idle time between 
the big and LITTLE cores (20\% and ~13\%, respectively), which drives to the conclusion that the fast cores stall waiting for
completion of tasks executed on the LITTLE cores. This fact can be confirmed in the final stages of the Botlev-OmpSs trace.

The previous observations pave the road to a combination of scheduling policy and execution model for AMPs, 
in which asymmetry is exploited through {\em ad-hoc} scheduling policies during the first stages of the factorization 
--when the potential parallelism is massive--, and this is later replaced with the use 
of asymmetric-aware kernels and coarse-grain VCs
in the final stages of the execution, when the concurrency is scarce. Both approaches are not mutually exclusive, 
but complementary depending on the level of task concurrency available at a given execution stage.

\begin{table}
\centering
\caption{Percentage of time per worker thread in idle or running state for different runtime configurations for the Cholesky factorization 
({\tt n}~=~6,144, {\tt b}~=~448).
Note that {\sc wt 0} is the master thread, and thus is never idle; for it, the rest of the time till
100\% percentage is devoted to {\tt synchronization}, {\tt scheduling} and {\tt thread creation}. For the rest of threads, this amount of time is devoted to {\tt runtime overhead}.}
\label{tab:th_state}

\ra{1.2}
\ca{2pt}

\renewcommand{\fg}[1]{{#1}} 
\renewcommand{\br}[1]{{#1}} 

{\scriptsize
\begin{tabular}{crrrrrrrrr} 
   	\toprule
                 & \phantom{a} & \multicolumn{2}{c}{OmpSs - Seq. BLIS} & \phantom{ab} & \multicolumn{2}{c}{OmpSs - Asym. BLIS} & \phantom{ab} & \multicolumn{2}{c}{Botlev-OmpSs - Seq. BLIS} \\ 
                 & \phantom{a} & \multicolumn{2}{c}{(4 worker threads)} & \phantom{ab} & \multicolumn{2}{c}{(4 worker threads)} & \phantom{ab} & \multicolumn{2}{c}{(8 worker threads, 4+4)} \\ 
                                          \cmidrule{3-4}                                         \cmidrule{6-7}                                      \cmidrule{9-10}
                       & \phantom{a} &    {\tt idle}& {\tt running}& \phantom{ab}  & {\tt idle}& {\tt running}& \phantom{ab} & {\tt idle}  & {\tt running} \\ \hline 
	 {\sc wt 0}    & \phantom{a} &    \fg{--}   & \fg{98.41}   & \phantom{ab}  & \fg{--}   & \fg{97.85}   & \phantom{ab} & \fg{--}     & \fg{86.53}    \\ \cline{3-10}
	 {\sc wt 1}    & \phantom{a} &    \br{5.59} & \fg{94.22}   & \phantom{ab}  & \fg{5.51} & \fg{94.29}   & \phantom{ab} & \fg{13.63}  & \fg{86.28}    \\ \cline{3-10}
	 {\sc wt 2}    & \phantom{a} &    \br{3.14} & \fg{96.67}   & \phantom{ab}  & \fg{5.27} & \fg{94.53}   & \phantom{ab} & \fg{13.94}  & \fg{85.98}    \\ \cline{3-10}
	 {\sc wt 3}    & \phantom{a} &    \br{5.77} & \fg{94.07}   & \phantom{ab}  & \fg{5.17} & \fg{94.62}   & \phantom{ab} & \fg{13.43}  & \fg{86.47}    \\ \cline{3-10}
	 {\sc wt 4}    & \phantom{a} &    \br{--}   & \fg{--}      & \phantom{ab}  & \fg{--}   & \fg{--}      & \phantom{ab} & \fg{19.26}  & \fg{80.51}    \\ \cline{3-10}
	 {\sc wt 5}    & \phantom{a} &    \br{--}   & \fg{--}      & \phantom{ab}  & \fg{--}   & \fg{--}      & \phantom{ab} & \fg{21.12}  & \fg{78.69}    \\ \cline{3-10}
	 {\sc wt 6}    & \phantom{a} &    \br{--}   & \fg{--}      & \phantom{ab}  & \fg{--}   & \fg{--}      & \phantom{ab} & \fg{20.84}  & \fg{78.97}    \\ \cline{3-10}
	 {\sc wt 7}    & \phantom{a} &    \br{--}   & \fg{--}      & \phantom{ab}  & \fg{--}   & \fg{--}      & \phantom{ab} & \fg{20.09}  & \fg{79.70}    \\ \bottomrule
	 {\sc Avg.}     & \phantom{a} &    \br{4.84} & \fg{95.89}    & \phantom{ab} & \fg{5.32} & \fg{94.90}   & \phantom{ab} & \fg{17.47}  & \fg{82.89}    \\ \bottomrule
\end{tabular}
}

\end{table}

\section{Conclusions}
\label{sec:conclusions}

In this paper, we have addressed the problem of refactoring existing runtime task schedulers to exploit task-level
parallelism in novel AMPs, focusing on ARM big.LITTLE systems-on-chip. 
We have demonstrated that, for the specific domain of DLA, an approach 
that delegates the burden of dealing with asymmetry to the library (in our case, using an asymmetry-aware BLIS implementation),
does not require any revamp of existing task schedulers, and can deliver high performance.
This proposal paves the road towards reusing conventional runtime schedulers for SMPs (and all the associated
improvement techniques developed through the past few years), as the runtime only has a symmetric view of the hardware. 
Our experiments reveal that this solution is competitive and even improves 
the results obtained with an asymmetry-aware scheduler for DLA operations.


\section*{Acknowledgments}

The researchers from Universidad Complutense de Madrid were supported by project CICYT TIN2012-32180.
Enrique S. Quintana-Ort\'i was supported by projects CICYT TIN2011-23283 and TIN2014-53495-R as well as the EU project FP7 318793 ``EXA2GREEN''.

\section*{References}
\label{sect:bib}
\bibliographystyle{elsarticle-num}
\bibliography{enrique,energy,asymmetric}   

\end{document}